\documentclass[11pt]{article}

\usepackage[final]{acl}

\usepackage{times}
\usepackage{latexsym}

\usepackage[T1]{fontenc}

\usepackage[utf8]{inputenc}

\usepackage{microtype}

\usepackage{inconsolata}

\usepackage{graphicx}

\usepackage{amsmath}
\usepackage{booktabs}
\usepackage{multirow}
\usepackage{tcolorbox}
\usepackage[T1]{fontenc}
\usepackage{xcolor}
\usepackage{subcaption}

\title{Improving Large Audio-Language Model Embedding Baselines for Instruction-Aware Speech Retrieval}

\author{Chen-An Li$^1$ \quad Cheng-Han Chiang$^1$ \quad  Hung-yi Lee$^{1,2}$ \\
$^1$National Taiwan University \\
$^2$NTU Artificial Intelligence Center of Research Excellence (NTU AI-CoRE) \\
\texttt{f13942069@ntu.edu.tw} \quad \texttt{dcml0714@gmail.com} \quad \texttt{hungyilee@ntu.edu.tw}
}

\begin{document}
\maketitle
\begin{abstract}
Speech retrieval finds relevant spoken documents given a spoken query.
The spoken query alone may not reveal the retrieval goal, such as finding an answer or matching speaker identity.
Instruction-aware speech retrieval expresses this goal through a natural-language text instruction.
Large audio-language models (LALMs) effectively process speech and text, and our analysis shows that different retrieval requirements favor hidden states from different audio-encoder layers.
We introduce \textbf{IRIS-Embed}, which adapts LALMs into embedding models through contrastive training and instruction-aware audio layer weighting.
We construct about 900K training samples from real data and synthetic data, with instructions expressing requirements for one or more speech attributes.
On INSPIRE, an instruction-aware speech retrieval benchmark, IRIS-Embed consistently outperforms the corresponding LALM baselines.
The best model achieves 62.92\% macro-average Recall@50, surpassing the strongest baseline at 45.15\%.
Controlled comparisons demonstrate instruction awareness and the benefits of audio layer weighting.
On MAEB, a massive audio embedding benchmark, macro-average gains of 9.4--19.8 points over the corresponding baselines show generalization beyond speech retrieval.
The code and data are publicly available at \url{https://github.com/lca0503/IRIS}.
\end{abstract}

\section{Introduction}
\label{sec:introduction}

Speech retrieval searches a collection of spoken documents using a spoken query~\citep{browsing,spokencontent,spokensurvey}.
For example, ``Where should we go this weekend?'' spoken excitedly could be used to find activity suggestions, another recording by the same speaker, or speech with similar excitement.
The spoken query alone does not distinguish these goals.

\begin{figure}[hbtp]
    \centering    
    \includegraphics[width=0.91\linewidth]{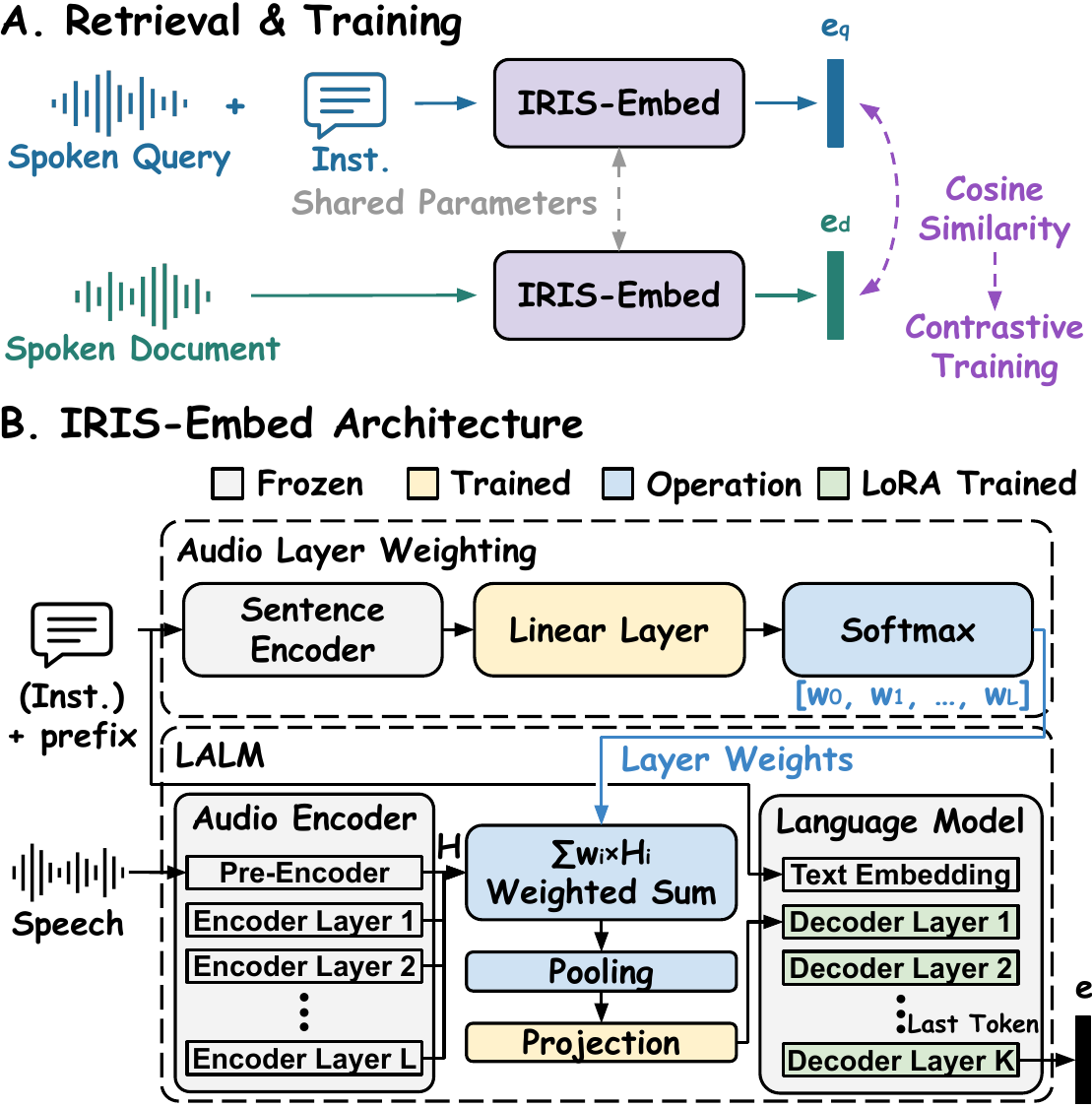}
    \caption{IRIS-Embed overview.
        A: Retrieval and training with a shared model.
        B: Audio layer weighting and embedding computation.
        Inst.: instruction.}
    \label{fig:model_framework}
    \vspace{-10pt}
\end{figure}

Instruction-aware speech retrieval uses natural-language instructions to express the user's retrieval goal~\citep{inspire}.
Each instruction expresses requirements for speech attributes such as spoken content, speaker identity, and speaking style.
The required values may come from the spoken query or be specified directly.
For example, ``Find speech by the same speaker, spoken happily'' asks for a document that matches the query's speaker identity and has a happy speaking style.
A relevant spoken document must satisfy every requirement.

Large audio-language models (LALMs) connect an audio encoder to a language model and have shown strong performance on tasks such as speech recognition, audio captioning, and audio question answering~\citep{qwen25omni,gemma4,audioflamingo3,desta25}.
Their ability to process speech and text within a model motivates their adaptation for instruction-aware speech retrieval. 
Our analysis shows that different retrieval requirements favor hidden states from different audio-encoder layers.
We ask: \emph{How can we adapt LALMs for instruction-aware speech retrieval?}

We introduce IRIS-Embed (\textbf{I}nstruction-aware \textbf{R}etr\textbf{I}eval for \textbf{S}peech), which adapts LALMs into embedding models through contrastive training and instruction-aware audio layer weighting. 
Figure~\ref{fig:model_framework} illustrates the overall framework.
IRIS-Embed separately encodes a spoken query with its instruction and each spoken document, then ranks documents by cosine similarity.
Audio-layer weighting predicts weights based on the instruction to combine hidden states from different layers of the frozen audio encoder before language-model encoding.

Contrastive training aligns query embeddings with relevant document embeddings.
We construct about 900K samples from real data and synthetic data, each pairing a spoken query and instruction with positive and negative spoken documents.
Each instruction expresses requirements for one or more speech attributes.
The positive spoken document must satisfy every requirement in the instruction.

We evaluate IRIS-Embed on INSPIRE~\citep{inspire}, an instruction-aware speech retrieval benchmark. 
Across two model families, IRIS-Embed consistently outperforms the corresponding LALM baselines, achieving macro-average Recall@50 scores of 62.92\% and 47.41\%, both surpassing the strongest baseline at 45.15\%.
Controlled comparisons demonstrate instruction awareness for single- and multi-attribute speech requirements.
We also show that audio-layer weighting improves speaker and speaking-style matching compared with using only the final audio-encoder hidden states.
On MAEB~\citep{maeb}, we evaluate IRIS-Embed on 15 diverse tasks, including gender classification, accent-pair classification, and artist retrieval. 
IRIS-Embed achieves macro-average gains of 9.4--19.8 points over the corresponding LALM baselines, demonstrating generalization beyond speech retrieval.

\section{Related Work}
\label{sec:related_work}

\subsection{Spoken Content Retrieval}
\label{subsec:spoken_content_retrieval}
Spoken content retrieval finds spoken documents with content relevant to a spoken query~\citep{trec, browsing,spokencontent,spokensurvey}.
Prior work explores query-by-example spoken term detection~\citep{query,cnnquery,neuralquery} and spoken passage retrieval for question answering~\citep{speechdpr,speechrag}.
We focus on instruction-aware speech retrieval, where instructions express requirements based on spoken content, speaker identity, speaking style, or other speech attributes.

\subsection{Instruction-Aware Retrieval}
\label{subsec:instruction_aware_retrieval}
Instructions guide text retrieval by expressing retrieval goals and relevance requirements~\citep{tart,instructor,promptriever,followir,e5mistral,nvembed,qwen3embedding,geminiembedding}.
Image and multimodal retrieval methods also use text to express retrieval goals~\citep{fashioniq,cirr,searle,pic2word,genecis,magiclens,e5v,umrb,megapairs,vlm2vec,vlm2vec2,qwen3vlembedding}.
INSPIRE~\citep{inspire} benchmarks instruction-aware speech retrieval with requirements for one or more speech attributes.
We adapts LALMs into embedding models for instruction-aware speech retrieval.

\subsection{Audio-Language Embeddings}
\label{subsec:al_embeddings}
Audio-language embeddings support retrieval between audio and text, as well as audio classification~\citep{audioclip,clap,laionclap,blat,wavcaps,cacophony,finelap}.
LALM embeddings also support retrieval between audio and text~\citep{lcoembedding,e5omni,omniembedaudio,alm2vec}.
We focus on retrieving spoken documents given a spoken query and a natural-language text instruction.
Furthermore, IRIS-Embed uses instruction-aware audio-layer weighting to combine hidden states from different layers of the frozen audio encoder before language-model encoding.

\subsection{Audio Layer Aggregation}
\label{subsec:audio_layer_aggregation}
SUPERB aggregates frozen self-supervised speech-encoder hidden states for downstream tasks~\citep{superb}.
Prompt-aware Mixture uses prompt-guided fusion for recognition and captioning~\citep{pam}.
WAVE and OmniUE aggregate language-model states for multimodal embeddings~\citep{wave,omniue}.
Instead, IRIS-Embed learns instruction-aware weights to combine hidden states from different layers of an LALM's frozen audio encoder through contrastive speech retrieval training.

\section{Method}
\label{sec:method}

\subsection{Problem Formulation}
\label{subsec:problem_formulation}

Instruction-aware speech retrieval ranks spoken documents in a collection $D$ given a spoken query $q$ and a natural-language text instruction $z$.
Speech attributes describe properties of a recording, such as spoken content, speaker identity, and speaking style.
The instruction $z$ expresses the requirements on these attributes by referring directly to the spoken query or specifying the values.
For example, ``Find speech by the same speaker, spoken happily'' requires a document to match $q$'s speaker identity and have a happy speaking style.
A relevant document must satisfy every requirement; unconstrained attributes do not affect relevance.

For each query--instruction pair $(q,z)$, let $D^{+} \subseteq D$ denote the positive documents relevant to $(q,z)$ and $D^{-} = D \setminus D^{+}$ denote the negative documents.
A scoring function $f(d \mid q,z)$ ranks documents $d \in D$ in decreasing order of score.
An ideal ranking satisfies $f(d^{+} \mid q,z) > f(d^{-} \mid q,z)$,
for every $d^{+} \in D^{+}$ and $d^{-} \in D^{-}$.

\subsection{Architecture}
\label{subsec:architecture}

As shown in Figure~\ref{fig:model_framework}, IRIS-Embed uses a shared model to separately encode the spoken query $q$ with instruction $z$ and each spoken document $d$.
For each recording, we extract hidden states from the pre-encoder and all layers of the frozen audio encoder.
The audio-layer weighting module encodes a text-only prompt to predict weights for combining these hidden states.
The weighted audio hidden states are then pooled and projected before being passed to the language model, which processes the text and weighted audio hidden states using its own prompt template.
Query prompts include instruction $z$ for both audio layer weighting and language-model encoding, whereas document prompts contain no instruction $z$.

We use the language model's final-layer hidden state of the last non-padding token as the embedding, obtaining $\mathbf{e}_q$ for the query--instruction pair $(q,z)$ and $\mathbf{e}_d$ for document $d$. 
Documents are ranked by decreasing cosine similarity, $f(d\mid q,z) = \cos\bigl(\mathbf{e}_q,\mathbf{e}_d\bigr)$.
We next explain the motivation and design of audio layer weighting.

\subsubsection{Audio Layer Weighting}
\label{subsubsec:audio_layer_weighting}

\begin{figure}[t]
    \centering
    \includegraphics[width=0.92\linewidth]{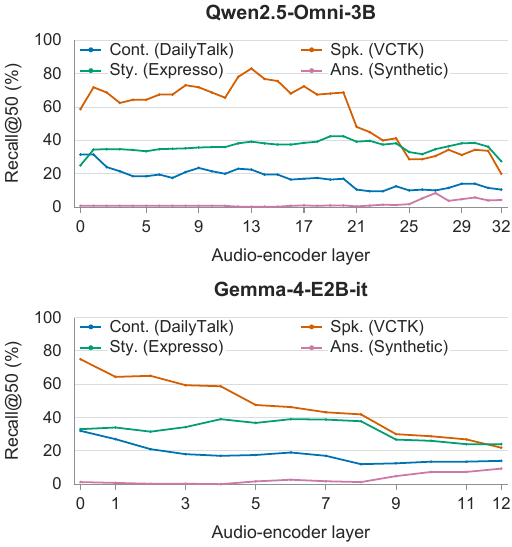}
    \caption{Recall@50 across frozen audio-encoder layers. 
        Layer 0 represents pre-encoder hidden states; layers 1 and above represent encoder hidden states.}
    \vspace{-8pt}
    \label{fig:layer_recall}
\end{figure}

To investigate how different audio-encoder layers perform across retrieval tasks, we evaluate their frozen hidden states on INSPIRE~\citep{inspire}.
For each layer, we mean-pool the hidden states over time and rank spoken documents by cosine similarity to the spoken query, without using instructions, the audio projection, or the language model.
Figure~\ref{fig:layer_recall} reports Recall@50 for dialogue continuation, speaker matching, speaking-style matching, and answer retrieval.
The best-performing layer varies across tasks, and the final layer does not consistently achieve the highest recall.
These findings motivate combining hidden states from multiple audio-encoder layers before passing them to the language model.

For each recording, let $\mathbf{H}_0$ denote the pre-encoder hidden states and $\mathbf{H}_1,\ldots,\mathbf{H}_L$ the hidden states produced by the $L$ audio-encoder layers.
We combine hidden states at the same time step to obtain weighted audio hidden states:
\begingroup
\setlength{\abovedisplayskip}{4pt}
\setlength{\belowdisplayskip}{4pt}
\setlength{\abovedisplayshortskip}{2pt}
\setlength{\belowdisplayshortskip}{2pt}
\begin{equation*}
    \widetilde{\mathbf{H}} = \sum_{\ell=0}^{L} w_\ell\mathbf{H}_\ell,
\end{equation*}
\endgroup
where $w_\ell$ is the scalar weight assigned to $\mathbf{H}_\ell$.

A shared audio layer weighting module computes weights from a text prompt $p$:
\begingroup
\setlength{\abovedisplayskip}{4pt}
\setlength{\belowdisplayskip}{4pt}
\setlength{\abovedisplayshortskip}{2pt}
\setlength{\belowdisplayshortskip}{2pt}
\begin{equation*}
    w_\ell = [\operatorname{softmax}(g(p))]_\ell,
\end{equation*}
\endgroup
where $g$ uses a pretrained sentence encoder followed by a linear layer to produce $L+1$ layer scores.
The notation $[\cdot]_\ell$ selects the weight for $\mathbf{H}_\ell$.
The resulting weights are nonnegative and sum to one.
For queries, the prompt $p$ includes the instruction $z$, allowing the layer weights to vary across retrieval tasks.
For documents, we use a fixed prompt, so all documents share the same layer weights.

\subsection{Training Objective}
\label{subsec:objective}
Each batch consists of $B$ samples $(q_i,z_i,d_i^{+},d_i^{-})$, where $i$ indexes samples and $d_i^{+}$ and $d_i^{-}$ are the paired positive and negative documents.
The batch-level document pool is $\mathcal{C}=\bigcup_{i=1}^{B}\{d_i^{+},d_i^{-}\}$.
Since documents from other samples may also be relevant to $(q_i,z_i)$, treating them as negatives could introduce false negatives.
We use training labels to mask these additional relevant documents.
For each sample $i$, we construct a document pool $\mathcal{C}_i$ by excluding such documents from $\mathcal{C}$ while preserving the paired positive document $d_i^{+}$.

We optimize the model using the InfoNCE loss~\citep{infonce}:
\begingroup
\setlength{\abovedisplayskip}{4pt}
\setlength{\belowdisplayskip}{4pt}
\setlength{\abovedisplayshortskip}{2pt}
\setlength{\belowdisplayshortskip}{2pt}
\begin{equation*}
    \mathcal{L}
    = -\frac{1}{B}\sum_{i=1}^{B}
    \log\frac{\exp\bigl(f(d_i^{+}\mid q_i,z_i)/\tau\bigr)}
    {\sum_{d\in\mathcal{C}_i}\exp\bigl(f(d\mid q_i,z_i)/\tau\bigr)},
\end{equation*}
\endgroup
where $\tau>0$ is a temperature parameter controlling the sharpness of the score distribution.
This objective encourages the model to assign higher relevance scores to the paired positive document than to the remaining negatives in $\mathcal{C}_i$.

During training, we freeze the audio encoder and the sentence encoder in $g$.
We adapt the language model using low-rank adaptation (LoRA)~\citep{lora}.
The audio projection, which maps audio hidden states to the language model's embedding dimension, is trained jointly with the added linear layer in $g$ and the LoRA parameters.

\section{Training Data}
\label{sec:training_data}

\subsection{Notation}
\label{subsec:data_notation}

We represent each recording $x$ by a set $A(x)$ of attribute-value pairs.
Each pair identifies a speech attribute and its value, such as $(\mathrm{style},\mathrm{happy})$ for a happy speaking style.

The requirements $r$ form a set of attribute-value pairs that a relevant spoken document must satisfy.
We write $r = r_q \cup r_s$.
The requirements in $r_q$ match the spoken query's attribute values, with $\emptyset \ne r_q \subseteq A(q)$.
The set $r_s$ contains attribute-value pairs with specified values that need not match the spoken query.
We allow $r_s = \emptyset$.
The text instruction $z$ expresses $r$ in natural language, referring to the spoken query for $r_q$ and stating the values in $r_s$.
The two sets use disjoint attribute keys to avoid conflicting requirements, so $r_q \cap r_s = \emptyset$.

For example, consider a spoken query $q$ with happy speech and a dog barking in the background.
The requirements $r = \{(\mathrm{environment},\mathrm{dog\ bark}), (\mathrm{style},\mathrm{sad})\}$ ask for sad speech with a barking dog.
Here, $r_q = \{(\mathrm{environment},\mathrm{dog\ bark})\}$ matches the query's environmental sound, while $r_s = \{(\mathrm{style},\mathrm{sad})\}$ specifies a sad speaking style.
A corresponding instruction $z$ is ``\textit{Find a recording with the same background sound as the query, spoken sadly.}''

\subsection{Data Format}
\label{subsec:data_format}

Each training sample $(q,z,d^{+},d^{-})$ contains a spoken query, an instruction, and a positive and a negative spoken document.
A document is relevant if it satisfies all requirements in $r$: $D^{+} = \{d \in D : r \subseteq A(d)\}$.
The paired positive belongs to $D^{+}$, while the paired negative belongs to $D^{-}$ and violates at least one requirement.
Attributes outside $r$ do not affect relevance.

\subsection{Construction}
\label{subsec:data_construction}
\begin{figure}[t]
    \centering    
    \includegraphics[width=0.92\linewidth]{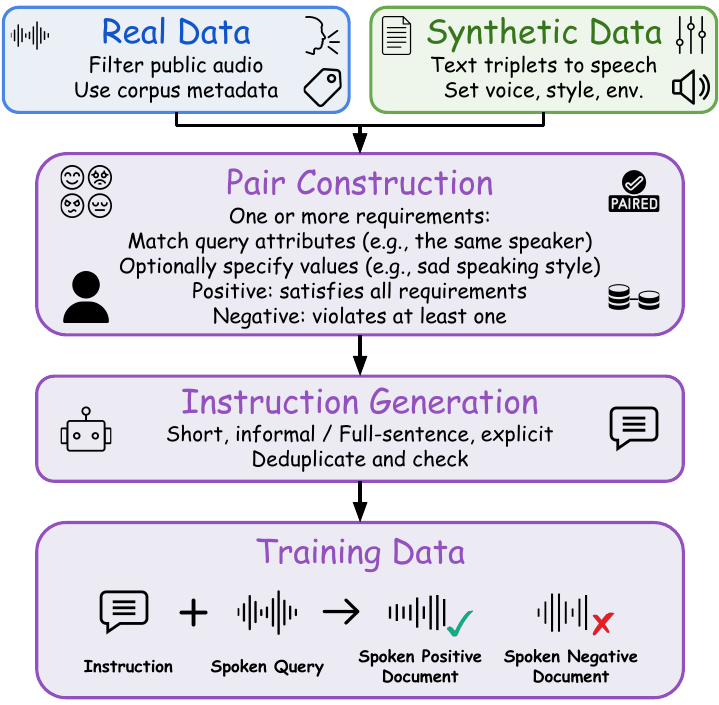}
    \caption{Overview of the data construction pipeline, combining real and synthetic data to create spoken query--document pairs and retrieval instructions.}
    \vspace{-8pt}
    \label{fig:data_framework}
\end{figure}

We construct IRIS-Embed training samples from real data and synthetic data, each paired with an instruction expressing its retrieval requirements.
Figure~\ref{fig:data_framework} shows the data construction pipeline.

\paragraph{Real Data.}
We use public speech and audio corpora.
The source annotations cover speaking style~\citep{cremad}, speaker identity~\citep{voxceleb}, gender~\citep{esd}, accent~\citep{l2arctic}, sentiment~\citep{meld}, dialogue~\citep{spokenwoz,multidialog}, and environmental sound~\citep{urbansound8k}.
We remove uninformative labels such as \texttt{unknown} and \texttt{other}, recordings outside 1--120~s, and environmental sounds containing speech.
We also map equivalent labels across corpora, such as \texttt{happy} and \texttt{joyful}, to a common value.
Corpus metadata provide attribute values for each recording, which we use to construct retrieval requirements and select query--document pairs.

\paragraph{Synthetic Data.}
We sample English text triplets from the MS MARCO~\citep{msmarco} hard-triplet data, each containing a query, a relevant passage, and a hard negative passage.
We synthesize speech from these texts using multiple text-to-speech models~\cite{gemini25,qwen3tts}.
Voice selection controls speaker identity, synthesis prompts control speaking style, and sound mixing adds environmental sounds.

\paragraph{Pair Construction.}
For each spoken query $q$, we sample a nonempty $r_q$ from $A(q)$ and optionally form $r_s$ by sampling values for other attribute keys annotated for $q$.
Speaker identity, dialogue, and answer relevance are restricted to $r_q$.
When speaker identity is included, we exclude gender and accent, since matching the speaker also guarantees matching these attributes in our data.
For each $r$, we sample a positive and a negative document, prioritizing negatives that violate fewer requirements to obtain harder negatives.

\paragraph{Instruction Generation.}
We use commercial large language models to express each set of requirements $r$ as a natural-language instruction $z$.
To cover diverse instruction styles, we use one prompt to generate short instructions and another to generate full-sentence instructions.
We remove near-duplicates using ROUGE-L~\citep{rouge}.
We also verify that each instruction covers all requirements in $r$, refers to the spoken query for pairs in $r_q$, and specifies the values in $r_s$.

\paragraph{Statistics.}
The resulting dataset contains approximately 900K samples: 800K from real data and 100K from synthetic data.
We use about 890K samples for training and 10K for validation.
Appendix~\ref{app:data} lists the sources, counts, and data construction settings.
\section{Experiments}
\label{subsec:experiment}

\subsection{Experimental Setup}
\label{subsec:experimental_setup}

\paragraph{Evaluation.}
We evaluate IRIS-Embed on four INSPIRE~\citep{inspire} subsets. 
DailyTalk~\citep{dailytalk} evaluates dialogue continuation, VCTK~\citep{vctk} evaluates speaker matching, and Expresso~\citep{expresso} evaluates speaker and speaking-style matching.
Speaking style includes emotion and vocal states.
The Synthetic subset evaluates answer retrieval and matching of speaker identity, speaking style, and environmental sounds, with instructions involving one or more attributes. 
Performance is measured by Recall@$K$ for $K\in\{10,50,100\}$, defined as the fraction of relevant documents retrieved in the top $K$ results, averaged across query--instruction pairs and reported as percentages.

\paragraph{Training.}
We train IRIS-Embed with Qwen2.5-Omni-3B~\citep{qwen25omni} and Gemma-4-E2B-it~\citep{gemma4} on the data in Section~\ref{sec:training_data}.
The audio layer weighting module contains a sentence encoder and a learned linear layer.
We use the frozen all-MiniLM-L6-v2 model~\citep{minilm,sbert} as the sentence encoder.
Training uses a global batch size of 512.
Appendix~\ref{app:training} provides further training details.

\begin{table*}[hbtp]
    \centering
    \small
    \setlength{\tabcolsep}{2.5pt}

    \resizebox{\linewidth}{!}{%
        \begin{tabular}{@{}ll|*{12}{r}|*{3}{r}@{}}
            \toprule

            \multirow{2}{*}{\textbf{Method}} & \multirow{2}{*}{\textbf{Model}}
            & \multicolumn{3}{c}{\textbf{DailyTalk}}
            & \multicolumn{3}{c}{\textbf{VCTK}}
            & \multicolumn{3}{c}{\textbf{Expresso}}
            & \multicolumn{3}{c|}{\textbf{Synthetic}}
            & \multicolumn{3}{c}{\textbf{Avg.}}
            \\
            \cmidrule(lr){3-5}\cmidrule(lr){6-8}\cmidrule(lr){9-11}\cmidrule(lr){12-14}\cmidrule(l){15-17}

            &
            & \textbf{R@10} & \textbf{R@50} & \textbf{R@100}
            & \textbf{R@10} & \textbf{R@50} & \textbf{R@100}
            & \textbf{R@10} & \textbf{R@50} & \textbf{R@100}
            & \textbf{R@10} & \textbf{R@50} & \textbf{R@100}
            & \textbf{R@10} & \textbf{R@50} & \textbf{R@100}
            \\

            \midrule

            RAND & Random
            & 0.24 & 1.06 & 2.10
            & 0.35 & 1.84 & 3.63
            & 0.30 & 1.50 & 2.96
            & 0.20 & 1.00 & 1.98
            & 0.27 & 1.35 & 2.67
            \\

            \midrule

            ENC & WavLM-large
            & 16.50 & 24.50 & 35.50
            & 17.50 & 35.00 & 43.75
            & 9.50 & 17.69 & 24.00
            & 1.70 & 4.48 & 7.22
            & 11.30 & 20.42 & 27.62
            \\

            \midrule

            \multirow{2}{*}{SENC} & ECAPA-TDNN
            & 15.50 & 24.00 & 29.50
            & \textbf{100.00} & \textbf{100.00} & \textbf{100.00}
            & \textbf{34.88} & 41.75 & 45.44
            & 10.02 & 14.83 & 18.82
            & \underline{40.10} & 45.15 & 48.44
            \\

            & emotion2vec-base
            & 11.50 & 23.00 & 30.50
            & 3.13 & 6.25 & 13.75
            & 0.88 & 4.38 & 8.00
            & 1.61 & 4.53 & 6.82
            & 4.28 & 9.54 & 14.77
            \\

            \midrule

            CAS & Qwen3-Embedding-8B
            & \textbf{62.00} & \textbf{81.50} & \textbf{89.50}
            & 1.88 & 6.25 & 10.63
            & 0.75 & 5.56 & 9.69
            & 7.02 & 8.95 & 11.07
            & 17.91 & 25.57 & 30.22
            \\

            \midrule

            \multirow{4}{*}{LALM} & Qwen2.5-Omni-3B
            & 26.50 & 53.00 & 64.00
            & 1.88 & 3.13 & 3.13
            & 0.75 & 3.75 & 6.69
            & 3.14 & 5.20 & 6.88
            & 8.07 & 16.27 & 20.18
            \\

            & Gemma-4-E2B-it
            & 9.00 & 25.50 & 33.50
            & 1.25 & 4.38 & 5.63
            & 0.13 & 1.19 & 2.25
            & 2.52 & 4.56 & 6.03
            & 3.23 & 8.91 & 11.85
            \\

            & Qwen3-Omni-30B
            & 51.50 & 74.50 & 83.00
            & 1.25 & 3.13 & 6.25
            & 0.94 & 3.56 & 6.19
            & 4.88 & 6.38 & 7.88
            & 14.64 & 21.89 & 25.83
            \\

            & LCO-Embedding-Omni-3B
            & 54.00 & 75.50 & 83.50
            & 1.25 & 2.50 & 5.63
            & 1.94 & 8.44 & 13.81
            & 6.82 & 9.37 & 11.65
            & 16.00 & 23.95 & 28.65
            \\

            \midrule

            \multirow{2}{*}{IRIS} & Qwen2.5-Omni-3B
            & \underline{60.00} & \underline{77.50} & \underline{87.50}
            & \underline{67.50} & \underline{85.00} & \underline{89.38}
            & \underline{29.63} & \textbf{45.00} & \textbf{53.88}
            & \textbf{26.06} & \textbf{44.17} & \textbf{54.78}
            & \textbf{45.80} & \textbf{62.92} & \textbf{71.39}
            \\

            & Gemma-4-E2B-it
            & 25.50 & 44.50 & 54.50
            & 58.75 & 74.38 & 79.38
            & 28.63 & \underline{43.81} & \underline{52.31}
            & \underline{15.07} & \underline{26.94} & \underline{36.08}
            & 31.99 & \underline{47.41} & \underline{55.57}
            \\

            \bottomrule
        \end{tabular}%
    }

    \caption{
        INSPIRE Recall@$K$.
        Avg.: macro-average across the four subsets.
        Bold / underline: best / second best.
        RAND: random; ENC / SENC: general / specialized encoders; CAS: cascaded retrieval; LALM: large audio--language models; IRIS: IRIS-Embed.
    }
    \label{tab:main_results}
\end{table*}

\begin{figure*}[hbtp]
    \centering    
    \includegraphics[width=\linewidth]{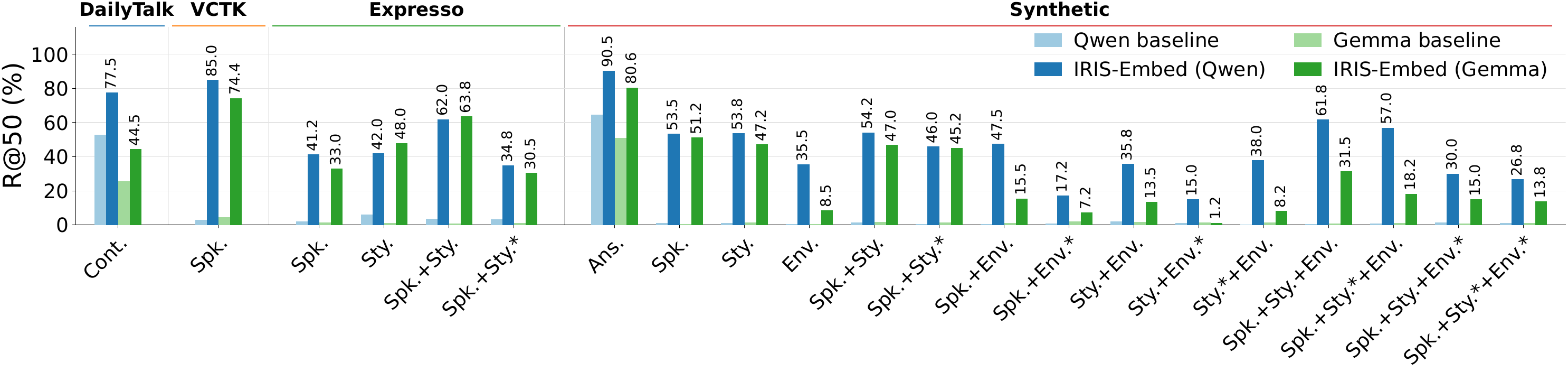}
    \caption{Recall@50 by instruction type on INSPIRE.
        Cont./Ans.: dialogue continuation/answer retrieval; Spk./Sty./Env.: speaker/speaking style/environmental sound.
        $+$: multiple attributes; $*$: specified values.}
    \label{fig:types}
\end{figure*}

\subsection{Comparison Methods}
\label{subsec:comparison_methods}

\paragraph{Random.}
Uniform random document rankings serve as a chance-level baseline~\citep{inspire}.

\paragraph{General Speech Encoders.}
We evaluate general speech representations using the self-supervised WavLM-large~\citep{wavlm} without instructions.
Query and document embeddings are obtained by mean-pooling its final-layer hidden states over time.

\paragraph{Specialized Speech Encoders.}
We evaluate specialized speech representations without instructions, using ECAPA-TDNN~\citep{ecapa_tdnn} for speaker embeddings and emotion2vec-base~\citep{emotion2vec} for emotion embeddings.

\paragraph{Cascaded Retrieval.}
We use a cascaded pipeline to retrieve spoken documents through text representations.
Whisper-large-v3~\citep{whisper} generates transcripts, while Qwen3-Omni-30B-A3B-Captioner~\citep{qwen3omni} generates captions.
Qwen3-Embedding-8B~\citep{qwen3embedding} encodes the transcripts and captions of each spoken query and spoken document, with the instruction included in the query input.

\paragraph{LALM Embeddings.}
We evaluate Qwen2.5-Omni-3B~\citep{qwen25omni}, Gemma-4-E2B-it~\citep{gemma4}, and Qwen3-Omni-30B~\citep{qwen3omni} without additional embedding training.
Embeddings are extracted from the final-layer hidden state of the last non-padding input token without generating a response.
We also include LCO-Embedding-Omni-3B~\citep{lcoembedding}, a Qwen2.5-Omni-based model trained for multimodal embeddings.

\subsection{Main Results}
\label{subsec:main_results}

Table~\ref{tab:main_results} shows that IRIS-Embed consistently outperforms the corresponding LALM baselines on all four INSPIRE subsets.
Macro-average Recall@50 reaches 62.92\% for Qwen2.5-Omni-3B and 47.41\% for Gemma-4-E2B-it, compared with 16.27\% and 8.91\% for their base models.
Both exceed ECAPA-TDNN's 45.15\%, the highest macro-average Recall@50 among comparison baselines.

Both IRIS-Embed models consistently outperform all LALM embedding baselines on VCTK and Expresso.
IRIS-Embed with Qwen2.5-Omni-3B achieves 85.00\% and 45.00\% Recall@50 on these subsets, respectively, compared with 2.50\% and 8.44\% for LCO-Embedding-Omni-3B.

Both models also consistently outperform all comparison baselines on Synthetic.
This subset includes instructions with multiple attributes, for which a relevant document must satisfy every requirement.
Recall@50 reaches 44.17\% for Qwen2.5-Omni-3B and 26.94\% for Gemma-4-E2B-it, compared with 14.83\% for ECAPA-TDNN.

On DailyTalk, IRIS-Embed with Qwen2.5-Omni-3B achieves 77.50\% Recall@50, close to 81.50\% for the cascaded pipeline using larger captioning and text-embedding models.
While the specialized ECAPA-TDNN remains strongest for speaker matching on VCTK, IRIS-Embed supports retrieval across speech attributes within a single embedding model.
\section{Analysis}
\label{sec:analysis}

\begin{figure*}[hbtp]
    \centering
    \includegraphics[width=0.92\linewidth]{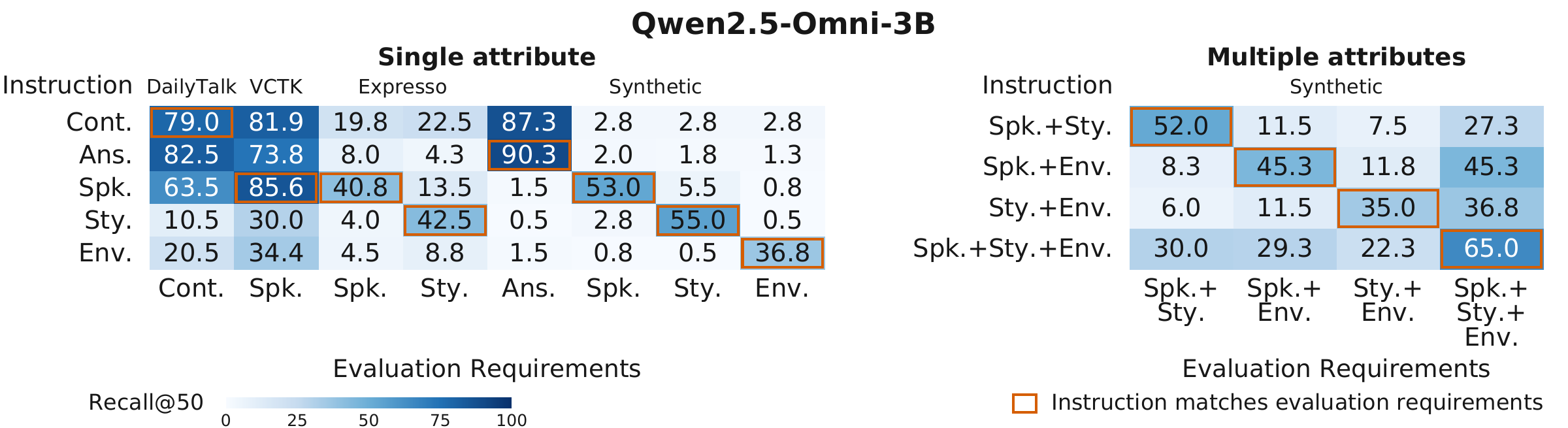}
    \caption{Recall@50 for IRIS-Embed with Qwen2.5-Omni-3B.
        Cont./Ans.: dialogue continuation/answer retrieval; Spk./Sty./Env.: speaker/speaking style/environmental sound; $+$: multiple attributes.}
    \label{fig:instruction_awareness}
    \vspace{-8pt}
\end{figure*}

\subsection{Instruction Types}
\label{subsec:instruction_types}

We breakdown IRIS-Embed's performance on instructions involving single or multiple attributes. 
Figure~\ref{fig:types} compares IRIS-Embed with the corresponding LALM baselines across instruction types. 
Both IRIS-Embed models achieve higher Recall@50 on most instruction types.

On Synthetic, IRIS-Embed also performs well on instructions involving multiple attributes, including specified attribute values. 
For joint matching of speaker identity, speaking style, and environmental sound, Recall@50 reaches 61.8\% for Qwen2.5-Omni-3B and 31.5\% for Gemma-4-E2B-it. 
For the speaker matching with a specified speaking style, the scores reach 46.0\% and 45.2\%, respectively. 
Both models outperform all baselines on most multi-attribute instruction types. 
The complete results are provided in the Appendix~\ref{app:full_results}.

\subsection{Instruction Awareness}
\label{subsec:instruction_awareness}
We examine whether IRIS-Embed distinguishes different retrieval goals for the same spoken query.
Figure~\ref{fig:instruction_awareness} reports Recall@50 for Qwen2.5-Omni-3B under different short instructions, such as ``Same speaker'' and ``Find the answer''.
Among five instructions with a single attribute, the instruction matching the evaluation requirements achieves the highest Recall@50 in seven of eight evaluation settings.
Switching from a speaker-matching instruction to an answer-retrieval instruction increases answer retrieval Recall@50 from 1.5\% to 90.3\%.

On Synthetic, we compare four short instructions combining speaker, speaking style, and environmental sound, such as ``Same speaker and same speaking style''.
The matching instruction achieves the highest Recall@50 in all four evaluation settings.
Gemma-4-E2B-it shows a similar pattern, and both models also follow full-sentence instructions, as reported in Appendix~\ref{app:instruction_awareness}.
These results show that IRIS-Embed follows the speech-attribute requirements expressed by the instruction.

\begin{table}[t]
    \centering
    \footnotesize
    \setlength{\tabcolsep}{3pt}

    \resizebox{\linewidth}{!}{
        \begin{tabular}{@{}l|rrrr|r@{}}
            \toprule
            \textbf{Method}
            & \textbf{DailyTalk}
            & \textbf{VCTK}
            & \textbf{Expresso}
            & \textbf{Synthetic}
            & \textbf{Avg.} \\
            \midrule
            Layer 13 & 15.50 & 76.88 & \textbf{48.94} & 35.27 & 44.15 \\
            Final Layer & \textbf{82.50} & 48.75 & 28.25 & 34.72 & 48.56 \\
            Mean Layer & 70.00 & 71.25 & 43.13 & 44.90 & 57.32 \\
            Finetuned & 32.50 & \textbf{91.25} & 47.38 & 43.35 & 53.62 \\
            Static Weight & 66.00 & 78.75 & 43.19 & \textbf{45.28} & 58.31 \\
            Inst. Weight (Ours) & 77.50 & 85.00 & 45.00 & 44.17 & \textbf{62.92} \\
            \bottomrule
        \end{tabular}
    }
    \caption{Recall@50 for Qwen2.5-Omni-3B audio layer aggregation on INSPIRE.
    Avg.: macro-average across the four subsets.
    }
    \vspace{-8pt}
    \label{tab:layer}
\end{table}

\subsection{Layer Aggregation}
\label{subsec:layer_aggregation}

Table~\ref{tab:layer} compares Qwen2.5-Omni-3B variants trained and evaluated with different audio-layer aggregation methods.
Layer 13 and Final Layer use the 13th and final audio-encoder hidden states, respectively.
Mean Layer uses the average weighting across layers.
Finetuned updates the audio encoder, while the other variants keep it frozen.
Static Weight learns a single set of audio layer weights shared across all queries and documents, regardless of instructions or prompts.
Inst. Weight is our instruction-aware audio layer weighting method, with weights varying according to the instruction.

Our method achieves the highest macro-average Recall@50, at 62.92\%.
Compared with Final Layer, our method improves recall on VCTK, Expresso, and Synthetic, increasing the macro-average by 14.36 points.
Layer 13 and Finetuned improve speaker and speaking-style matching over Final Layer, but have much lower dialogue-continuation recall.
These results suggest that retrieval tasks benefit from hidden states of different audio-encoder layers, highlighting the importance of instruction-aware audio layer weighting.

Although Mean Layer and Static Weight perform better on Synthetic, our method outperforms both on the other three subsets, achieving macro-average Recall@50 improvements of 5.60 and 4.61 points, respectively. 
These results demonstrate the effectiveness of instruction-aware weighting over instruction-independent layer aggregation.

With Gemma-4-E2B-it, our method also outperforms Final Layer on VCTK, Expresso, and Synthetic, showing improvements across multiple retrieval subsets.
Appendix~\ref{app:layer_aggregation} provides the full results for both models.
Together, these results demonstrate the effectiveness of instruction-aware audio layer weighting in improving retrieval performance while keeping the audio encoder frozen.

\begin{table*}[hbtp]
    \centering
    \small
    \setlength{\tabcolsep}{2pt}

    \resizebox{0.98\linewidth}{!}{%
        \begin{tabular}{@{}ll|*{15}{r}|r@{}}
            \toprule

            & & \multicolumn{9}{c}{Clf}
            & \multicolumn{2}{c}{Clust}
            & \multicolumn{2}{c}{PC}
            & \multicolumn{1}{c}{Rrnk}
            & \multicolumn{1}{c|}{Rtrvl}
            &
            \\
            \cmidrule(lr){3-11}\cmidrule(lr){12-13}\cmidrule(lr){14-15}\cmidrule(lr){16-16}\cmidrule(lr){17-17}

            \textbf{Method} & \textbf{Model}
            & \rotatebox{90}{\shortstack[l]{Beijing\\Opera}}
            & \rotatebox{90}{BirdCLEF}
            & \rotatebox{90}{\shortstack[l]{CommonLang\\Age}}
            & \rotatebox{90}{\shortstack[l]{GTZAN\\Genre}}
            & \rotatebox{90}{\shortstack[l]{IEMOCAP\\Gender}}
            & \rotatebox{90}{MInDS14}
            & \rotatebox{90}{\shortstack[l]{Mridingham\\Tonic}}
            & \rotatebox{90}{SIBFLEURS}
            & \rotatebox{90}{\shortstack[l]{VoxPopuli\\LanguageID}}
            & \rotatebox{90}{\shortstack[l]{Vehicle\\Sound}}
            & \rotatebox{90}{\shortstack[l]{VoxPopuli\\Gender}}
            & \rotatebox{90}{NMSQA}
            & \rotatebox{90}{\shortstack[l]{VoxPopuli\\Accent}}
            & \rotatebox{90}{GTZAN}
            & \rotatebox{90}{\shortstack[l]{JamAlt\\Artist}}
            & \textbf{Avg.}
            \\

            \midrule

            ENC & WavLM-large
            & 57.7 & 19.8 & 16.9 & 67.7 & 63.1 & 13.5 & 31.4 & 12.5 & 89.0
            & 0.2 & 0.1
            & 49.0 & 52.2
            & 68.8 & 78.5
            & 41.4
            \\

            \midrule

            \multirow{2}{*}{CLAP} & clap-htsat-fused
            & 92.8 & 18.0 & 15.3 & 67.9 & 93.6 & 9.2 & 47.0 & 9.1 & 32.0
            & 4.7 & 52.7
            & 47.9 & 53.2
            & 61.3 & 87.7
            & 46.2
            \\

            & larger\_clap\_general
            & 93.6 & 17.0 & \textbf{20.5} & 84.5 & 89.3 & 8.0
            & 49.4 & 9.4 & 84.6
            & 3.4 & 3.3
            & 47.4 & 53.5
            & 66.8 & \textbf{95.6}
            & 48.4
            \\

            \midrule

            \multirow{4}{*}{LALM} & Qwen2-Audio-7B
            & \textbf{97.4} & \textbf{37.1} & 17.6 & \textbf{93.1}
            & 93.0 & 24.4 & \textbf{61.2} & 15.4 & 99.0
            & 5.5 & 0.1
            & 48.9 & 52.9
            & \textbf{80.8} & 37.6
            & 50.9
            \\

            & LCO-Embedding-Omni-3B
            & 93.2 & 31.6 & 16.9 & 81.0 & 59.4 & \textbf{88.9}
            & 39.5 & \textbf{45.1} & 63.4
            & 2.2 & 1.0
            & \textbf{96.9} & 50.7
            & 75.4 & 67.1
            & \textbf{54.2}
            \\

            & Qwen2.5-Omni-3B$^{\dagger}$
            & 80.1 & 10.6 & 16.7 & 75.9 & 71.9 & 50.4 & 26.6 & 43.3 & 94.6
            & 2.3 & 1.2
            & 69.3 & 50.1
            & 56.6 & 11.4
            & 44.1
            \\

            & Gemma-4-E2B-it$^{\dagger}$
            & 63.6 & 4.0 & 15.5 & 28.2 & 50.5 & 22.0
            & 27.0 & 21.9 & \textbf{99.4}
            & 1.2 & 0.1
            & 57.2 & 50.2
            & 38.2 & 17.8
            & 33.1
            \\

            \midrule

            \multirow{2}{*}{IRIS} & Qwen2.5-Omni-3B
            & 91.9 & 13.3 & 14.3 & 80.1 & 98.7 & 24.4 & 33.5 & 20.9 & 71.0
            & 1.1 & 55.2
            & 92.9 & 61.5
            & 66.1 & 77.6
            & 53.5
            \\

            & Gemma-4-E2B-it
            & 88.5 & 15.1 & 16.4 & 60.6 & \textbf{99.0} & 14.2
            & 32.3 & 26.6 & 79.2
            & \textbf{8.2} & \textbf{68.5}
            & 94.3 & \textbf{63.2}
            & 60.6 & 67.0
            & 52.9
            \\

            \bottomrule
        \end{tabular}
    }

    \caption{
        Scores on 15 MAEB audio tasks.
        Avg.: macro-average across the 15 tasks.
        Higher is better; bold marks column maxima.
        Clf: classification; Clust: clustering; PC: pair classification; Rrnk: reranking; Rtrvl: retrieval.
        CLAP: contrastive language--audio pretraining.
        $^{\dagger}$: baselines using the same task instructions as IRIS-Embed.
    }
    \vspace{-8pt}
    \label{tab:maeb}
\end{table*}

\subsection{Layer Weights}
\label{subsec:layer_weights}

\begin{figure}[t]
    \centering
    \includegraphics[width=\linewidth]{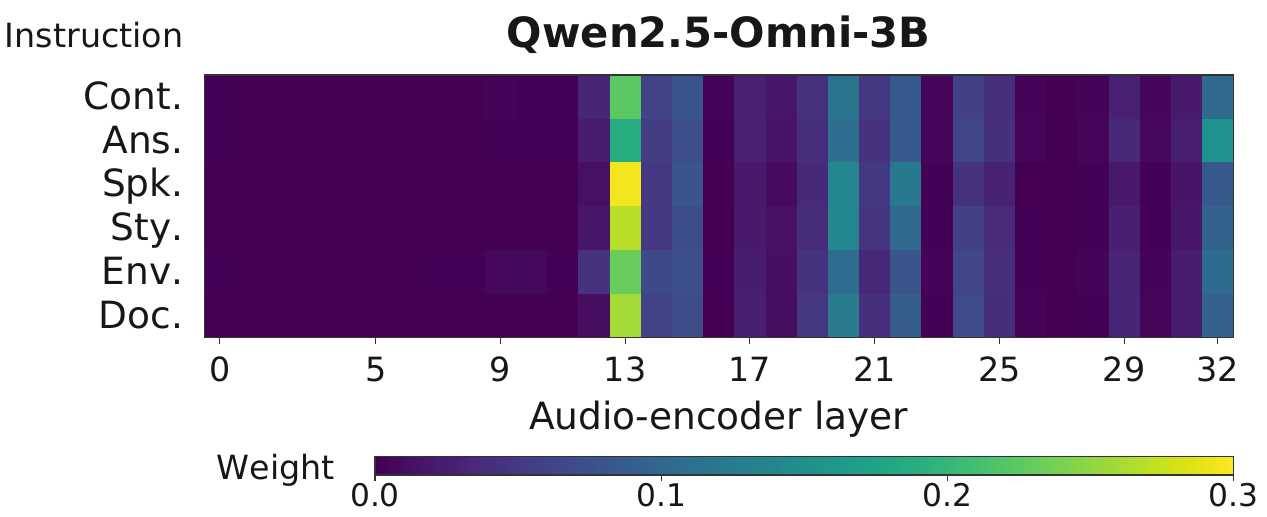}
    \caption{Scale-adjusted audio layer weights for Qwen2.5-Omni-3B. 
        Doc.: document prompt; layer 0: pre-encoder hidden states; encoder layers start at 1.}
    \vspace{-8pt}
    \label{fig:layer_weights}
\end{figure}

We analyze how audio layer weights vary across retrieval requirements.
Figure~\ref{fig:layer_weights} compares Qwen2.5-Omni-3B layer weights under five short instructions, such as ``Same speaker'' and ``Find the answer'', and the document prompt.
To account for scale differences, we adjust the displayed weights using the root mean square of each layer's hidden states.
This adjustment is used only for analysis.
Layer 13 receives the highest adjusted weight across all five query instructions.
Compared with speaker matching, answer retrieval places more weight on the final layer and less on layer 13.
Gemma-4-E2B-it shows a similar pattern, with answer retrieval shifting weight from layer 4 toward its final layer.
These weights show that instructions change the balance between intermediate and final audio-encoder hidden states.
Appendix~\ref{app:layer_weights} provides the full results for both models.

\subsection{Task Generalization}
\label{subsec:task_generalization}

We evaluate IRIS-Embed beyond retrieval tasks in INSPIRE on 15 MAEB~\citep{maeb} audio tasks covering classification, clustering, pair classification, reranking, and retrieval.
These include gender and accent tasks not covered by INSPIRE.
We encode each recording with a task instruction using the query prompt, except for retrieval documents, which use the document prompt without an instruction.
The models remain frozen during evaluation, and tasks derived from our retrieval training corpora are excluded.
The corresponding LALM baselines receive the same task instructions.
Appendix~\ref{app:maeb} provides evaluation details.

Table~\ref{tab:maeb} reports macro-average scores of 53.5 for Qwen2.5-Omni-3B IRIS-Embed and 52.9 for Gemma-4-E2B-it IRIS-Embed.
These exceed the corresponding LALM baselines scores by 9.4 and 19.8 points, respectively, and approach LCO-Embedding-Omni-3B's 54.2.
Both IRIS-Embed models outperform all comparison baselines on IEMOCAP gender classification, VoxPopuli gender clustering, and VoxPopuli accent pair classification.
These results show that IRIS-Embed generalizes beyond instruction-aware speech retrieval.
\section{Conclusion}
\label{sec:conclusion}
IRIS-Embed adapts LALMs into embedding models for instruction-aware speech retrieval through contrastive training and instruction-aware audio layer weighting.
Across two model families, macro-average Recall@50 on INSPIRE reaches 62.92\% and 47.41\%, compared with 45.15\% for the strongest baseline.
For Qwen2.5-Omni-3B, layer weighting improves VCTK Recall@50 from 48.75\% to 85.00\% over using only the final audio-encoder output.
Controlled comparisons show that the models follow instructions with a single attribute or multiple attributes.
On 15 MAEB tasks, macro-average scores improve by 9.4--19.8 points over the corresponding LALM baselines, showing generalization beyond speech retrieval.
These results show that IRIS-Embed enables effective retrieval across speech attributes.

\section*{Limitations}
Despite the advances achieved by our method, several limitations remain.

We prioritize smaller embedding models for computational efficiency and train IRIS-Embed with only two LALM backbones. 
Thus, its scalability to larger models remains unexplored.

IRIS-Embed combines hidden states from frozen pretrained audio encoders, making it dependent on the information they preserve. 
We do not explore alternative or complementary encoders that may capture broader speech attributes. 
Moreover, layer-weight visualizations do not establish individual layers' contributions to retrieval performance.

Our retrieval requirements rely on annotations covering limited speech attributes and values. 
Annotation and classifier errors may affect document-pair construction and in-batch masking, respectively. 
Our training also covers limited languages, voices, speaking styles, and environmental sounds, restricting its diversity relative to natural speech.

Finally, our cascaded baseline uses transcripts and audio captions. 
We leave comparisons with finer-grained cascaded approaches combining specialized speech embeddings or attribute-specific audio descriptions to future work.

While these limitations remain, IRIS-Embed demonstrates the effectiveness of adapting LALMs for instruction-aware speech retrieval. 
We believe our work will inspire further research on embeddings that follow diverse user intents and generalize across audio tasks.

\section*{Ethics Statement}
We use public audio corpora and synthetic data, with corpus licenses listed in Table~\ref{tab:datasets}.
Reuse should respect source licenses and speech-synthesis service terms, and shared synthetic recordings should be clearly labeled.

Speech-attribute matching can enable unauthorized speaker tracking or profiling.
Applications should obtain appropriate consent, protect recordings and embeddings, and avoid discriminatory uses.
Dataset biases may cause unequal retrieval quality, so deployment should include evaluation across speaker groups and recording conditions.

\section*{Acknowledgments}
We thank Yu-Xiang Lin from National Taiwan University for their helpful discussions and feedback. 
This work was financially supported by the National Science and Technology Council (NSTC) in Taiwan. 
We thank to National Center for High-performance Computing (NCHC) of National Applied Research Laboratories (NARLabs) in Taiwan for providing computational and storage resources. 
This work was supported by the Ministry of Education (MOE) of Taiwan under the project Taiwan Centers of Excellence in Artificial Intelligence, through the NTU Artificial Intelligence Center of Research Excellence (NTU AI-CoRE). 

\bibliography{custom}

\clearpage
\appendix
\raggedbottom

\section{Data Construction}
\label{app:data}

\subsection{Data Sources}
\label{app:data_sources}
\begin{table*}[t]
    \centering
    \small
    \setlength{\tabcolsep}{4pt}
    \begin{tabular}{lccccccccl}
        \toprule
        & \multicolumn{8}{c}{\textbf{Attribute}} & \\
        \cmidrule(lr){2-9}
        \textbf{Source}
        & \rotatebox{90}{\textbf{Answer relevance}}
        & \rotatebox{90}{\textbf{Speaking style}}
        & \rotatebox{90}{\textbf{Gender}}
        & \rotatebox{90}{\textbf{Speaker}}
        & \rotatebox{90}{\textbf{Accent}}
        & \rotatebox{90}{\textbf{Sentiment}}
        & \rotatebox{90}{\textbf{Dialogue}}
        & \rotatebox{90}{\textbf{Environmental sound}}
        & \textbf{License} \\
        \midrule
        \multicolumn{10}{l}{\textit{Real data}} \\
        VoxCeleb1~\citep{voxceleb} & & $\circ$ & $\bullet$ & $\bullet$ & & & & & CC BY-SA 4.0 \\
        ESD~\citep{esd} & & $\bullet$ & $\bullet$ & $\bullet$ & & & & & Custom \\
        MELD~\citep{meld} & & $\bullet$ & $\circ$ & $\bullet$ & & $\bullet$ & & & GPL-3.0 \\
        CREMA-D~\citep{cremad} & & $\bullet$ & $\bullet$ & & & & & & ODbL 1.0 \\
        L2-ARCTIC~\citep{l2arctic} & & $\circ$ & $\circ$ & $\bullet$ & $\bullet$ & & & & CC BY-NC 4.0 \\
        LibriSpeech-C~\citep{librispeechc} & & $\circ$ & $\circ$ & & & & & $\bullet$ & CC BY 4.0 \\
        EmoV-DB~\citep{emovdb} & & $\bullet$ & $\bullet$ & $\bullet$ & & & & & Custom \\
        MSP-IMPROV~\citep{mspimprov} & & $\bullet$ & $\bullet$ & & & & & & Custom \\
        VocalSound~\citep{vocalsound} & & & & & & & & $\bullet$ & CC BY-SA 4.0 \\
        SpokenWOZ~\citep{spokenwoz} & & & & & & & $\bullet$ & & CC BY-NC 4.0 \\
        MultiDialog~\citep{multidialog} & & & & & & & $\bullet$ & & CC \\
        UrbanSound8K~\citep{urbansound8k} & & & & & & & & $\bullet$ & CC BY-NC 3.0 \\
        EMNS~\citep{emns} & & $\bullet$ & $\circ$ & & & & & & Apache 2.0 \\
        \midrule
        \multicolumn{10}{l}{\textit{Synthetic data}} \\
        MS MARCO~\citep{msmarco} & $\bullet$ & $\bullet$ & & $\bullet$ & & & & $\bullet$ & CC BY 4.0 \\
        \bottomrule
    \end{tabular}
    \caption{
        Data sources and available annotations.
        $\bullet$ indicates corpus metadata or synthesis annotations used to construct retrieval requirements and select positive and negative spoken documents.
        $\circ$ indicates classifier predictions used only to mask additional relevant in-batch documents during training.
    }
    \label{tab:datasets}
\end{table*}

Table~\ref{tab:datasets} summarizes the data sources, available annotations, and licenses for real and synthetic data.
Speaking-style annotations include emotion and vocal states.

We use corpus metadata to construct retrieval requirements and select positive and negative spoken documents.
Classifier predictions supplement these annotations only to identify and mask additional relevant in-batch documents during training, as described in Section~\ref{subsec:objective}.

\subsection{Real Data}
\label{app:real_data}

\paragraph{Preprocessing.}
We remove uninformative labels such as \texttt{unknown} and \texttt{other}, along with recordings outside 1--120~s.
We exclude environmental-sound classes containing speech and map equivalent labels across corpora, such as \texttt{happy} and \texttt{joyful}, to a common value.

\paragraph{Classifiers.}
To supplement incomplete corpus annotations, we use audEERING's Wav2Vec2 age and gender model~\citep{agegender}\footnote{\url{https://huggingface.co/audeering/wav2vec2-large-robust-24-ft-age-gender}} to predict gender and emotion2vec+ large~\citep{emotion2vec}\footnote{\url{https://huggingface.co/emotion2vec/emotion2vec_plus_large}} to predict emotion.
For speakers with multiple recordings, we assign a shared predicted gender label through majority voting.
Predicted labels are used only for relevance checks during training.

\subsection{Synthetic Data}
\label{app:synthetic_data}
\begin{table*}[t]
    \centering
    \small
    \setlength{\tabcolsep}{4pt}
    \begin{tabular}{p{0.16\linewidth}p{0.80\linewidth}}
        \toprule
        \textbf{TTS model} & \textbf{Voice names} \\
        \midrule
        Gemini & Achernar, Achird, Algenib, Algieba, Alnilam, Aoede, Autonoe, Callirrhoe, Charon, Despina, Enceladus, Erinome, Fenrir, Gacrux, Iapetus, Kore, Laomedeia, Leda, Orus, Puck, Pulcherrima, Rasalgethi, Sadachbia, Sadaltager, Schedar, Sulafat, Umbriel, Vindemiatrix, Zephyr, Zubenelgenubi \\
        \addlinespace
        Qwen3 TTS & Ryan, Aiden \\
        \bottomrule
    \end{tabular}
    \caption{
        Synthesis voices used for synthetic data construction.
    }
    \label{tab:synthesis_voices}
\end{table*}
\begin{table*}[t]
    \centering
    \small
    \setlength{\tabcolsep}{4pt}
    \begin{tabular}{p{0.12\linewidth}p{0.84\linewidth}}
        \toprule
        \textbf{Style} & \textbf{Prompt} \\
        \midrule
        Happy &
        \textbf{Voice:} bright, cheerful, and upbeat, as if you are smiling while you speak.
        \textbf{Tone:} warm, joyful, and openly affectionate.
        \textbf{Pacing:} lively and energetic, with buoyant rhythm.
        \textbf{Emotion:} genuinely delighted and optimistic, emphasizing key words with an uplifted intonation. \\
        \addlinespace[3pt]
        Sad &
        \textbf{Voice:} soft, low, and melancholic, with a gentle tremor.
        \textbf{Tone:} heavy, sorrowful, and subdued, as if holding back tears.
        \textbf{Pacing:} slow and drawn out, with thoughtful pauses.
        \textbf{Emotion:} deep sadness and quiet grief, letting the weight of the words settle. \\
        \addlinespace[3pt]
        Angry &
        \textbf{Voice:} sharp, intense, and forceful, with hard consonants.
        \textbf{Tone:} confrontational, tense, and biting.
        \textbf{Pacing:} clipped and driving, with emphatic stress on charged words.
        \textbf{Emotion:} strong anger held just under control, barely restrained frustration. \\
        \addlinespace[3pt]
        Fearful &
        \textbf{Voice:} trembling, breathy, and slightly higher in pitch.
        \textbf{Tone:} anxious, uncertain, and vulnerable.
        \textbf{Pacing:} uneven, with nervous hesitations and quick shallow breaths.
        \textbf{Emotion:} palpable fear, as if something dangerous is nearby and you are trying not to be heard. \\
        \addlinespace[3pt]
        Surprised &
        \textbf{Voice:} wide-eyed and astonished, with sudden lifts in pitch.
        \textbf{Tone:} amazed and caught off guard.
        \textbf{Pacing:} quick bursts punctuated by short breathy intakes.
        \textbf{Emotion:} genuine shock and disbelief, emphasizing unexpected words with emphatic stress. \\
        \addlinespace[3pt]
        Disgusted &
        \textbf{Voice:} curled-lipped and repulsed, with nasal resonance.
        \textbf{Tone:} scornful, dismissive, and revolted.
        \textbf{Pacing:} slightly drawn-out on offensive words, with sharp cutoffs elsewhere.
        \textbf{Emotion:} strong aversion, as if the subject is physically distasteful. \\
        \addlinespace[3pt]
        Excited &
        \textbf{Voice:} thrilled, animated, and high-energy, with rising intonation.
        \textbf{Tone:} enthusiastic and eager, barely able to contain yourself.
        \textbf{Pacing:} fast and breathless, tumbling from one idea to the next.
        \textbf{Emotion:} exuberant anticipation and joyful hype. \\
        \addlinespace[3pt]
        Sleepy &
        \textbf{Voice:} drowsy, soft, and slightly slurred.
        \textbf{Tone:} lethargic, relaxed, and half-awake.
        \textbf{Pacing:} slow and drawn out, with yawning pauses.
        \textbf{Emotion:} deep tiredness, as if you are about to drift off mid-sentence. \\
        \addlinespace[3pt]
        Sarcastic &
        \textbf{Voice:} dry, flat, and ironic, with pointed emphasis.
        \textbf{Tone:} mocking and knowing, clearly meaning the opposite of the words.
        \textbf{Pacing:} deliberate, with exaggerated stress on key words.
        \textbf{Emotion:} wry amusement and condescending disbelief. \\
        \addlinespace[3pt]
        Whispering &
        \textbf{Voice:} quiet, breathy whisper with almost no vocal cord engagement.
        \textbf{Tone:} intimate, secretive, and conspiratorial.
        \textbf{Pacing:} careful and deliberate, with clear articulation despite the low volume.
        \textbf{Emotion:} hushed urgency, as if sharing a secret that must not be overheard. \\
        \bottomrule
    \end{tabular}
    \caption{
        Synthesis prompts for ten speaking styles, describing voice, tone, pacing, and emotion.
    }
    \label{tab:synthesis_styles}
\end{table*}

\paragraph{Text Selection.}
We sample 10,000 triplets from the MS MARCO~\citep{msmarco} hard-triplet dataset processed by Sentence-Transformers.\footnote{\url{https://huggingface.co/datasets/sentence-transformers/msmarco-bm25}}
Each triplet contains a query, a relevant passage, and a hard negative passage.
To reduce false negatives, the source dataset uses the \texttt{ms-marco-MiniLM-L6-v2} cross-encoder\footnote{\url{https://huggingface.co/cross-encoder/ms-marco-MiniLM-L6-v2}} to score the top 50 BM25-retrieved passages per query.
It retains negatives with scores more than 3.0 below the positive passage's score.
We retain English queries and passages of 5--150 tokens and remove near-duplicates using ROUGE-L~\citep{rouge} with a threshold of 0.7.

\paragraph{Speech Synthesis.}
We synthesize speech from these texts using Gemini~2.5~Pro TTS~\citep{gemini25}\footnote{\url{https://blog.google/innovation-and-ai/models-and-research/google-deepmind/gemini-2-5-native-audio/}}, Gemini~3.1~Flash TTS,\footnote{\url{https://blog.google/innovation-and-ai/models-and-research/gemini-models/gemini-3-1-flash-tts/}} and Qwen3 TTS~\citep{qwen3tts}.
Voice selection determines speaker identity.
The two Gemini models share 30 voices, while Qwen3 TTS uses Ryan and Aiden, as detailed in Table~\ref{tab:synthesis_voices}.

\paragraph{Speaking Styles.}
Synthesis prompts control ten speaking styles: happy, sad, angry, fearful, surprised, disgusted, excited, sleepy, sarcastic, and whispering.
Each prompt describes the voice, tone, pacing, and emotion.
Table~\ref{tab:synthesis_styles} provides the full prompts.

\paragraph{Environmental Sound.}
We use clips from eight UrbanSound8K~\citep{urbansound8k} classes: air conditioner, car horn, dog bark, drilling, engine idling, gunshot, jackhammer, and siren.
Before cropping, we discard clips with root-mean-square levels below $-45$~dB relative to full scale.
For each speech recording, we uniformly sample an environmental clip that passes this threshold.

\paragraph{Sound Mixing.}
We convert signals to mono by averaging channels and resample the environmental clip to match the speech sample rate.
If the clip is longer than the speech, we randomly crop a segment matching the speech duration.
Otherwise, we retain the full clip and place it at a uniformly sampled position within the speech.
We uniformly sample a signal-to-noise ratio from $\{5,10,15\}$~dB and scale the environmental signal using the mean squared amplitudes of both signals over the mixing interval.

\subsection{Pair Construction}
\label{app:data_pairs}

We construct retrieval requirements and document pairs from the annotations described in Appendix~\ref{app:data_sources}, following Section~\ref{subsec:data_format}.

We balance sampling across attribute keys and values in the retrieval requirements.
For each spoken query $q$, we sample a nonempty subset $r_q \subseteq A(q)$.
To form $r_s$, we may select attribute keys annotated for $q$ but not used in $r_q$.
For each selected key, we sample a value to be specified in the instruction.
Speaker identity, dialogue, and answer relevance appear only in $r_q$ and refer to the spoken query.
When $r$ includes speaker identity, we exclude gender and accent, since matching the speaker also guarantees matching these attributes in our data.

\paragraph{Pair Selection.}
For each spoken query and set of requirements $r$, we select a positive document satisfying all requirements and a negative document violating at least one.
We prioritize documents that violate fewer requirements in $r$ to obtain harder negatives.

\subsection{Instruction Generation}
\label{app:data_instructions}
\begin{table*}[t]
    \centering
    \small
    \setlength{\tabcolsep}{6pt}
    \begin{tabular}{p{0.47\linewidth}p{0.47\linewidth}}
        \toprule
        \textbf{Short instructions} & \textbf{Full-sentence instructions} \\
        \midrule
        \raggedright\ttfamily
        You are generating \{N\} very short, informal retrieval cues for an audio retrieval system.
        Given a spoken query, the user wants to find another clip satisfying certain requirements.
        \par\smallskip
        MATCHING constraints (same as spoken query):
        \par
        - the same \{key\} (e.g., \{values\}) as the given spoken query
        \par\smallskip
        TARGET constraints (find a specific value):
        \par
        - a specific \{key\} (e.g., \{values\}) of "\{value\}"
        \par\smallskip
        Each cue should be an informal, telegraphic phrase a user might type in a chat or voice interface: keyword-style fragments, not full sentences.
        \par\smallskip
        Requirements:
        \par
        1. Very short, typically 2 to 6 words
        \par
        2. Lowercase is fine, no trailing punctuation
        \par
        3. No full sentences; use fragments, comma-separated keywords, or elliptical phrases
        \par
        4. For MATCH constraints use phrases such as "same emotion", "matching accent"
        \par
        5. For TARGET constraints use phrases such as "female", "happy", "british accent"
        \par
        6. A natural example for this pattern is "\{example\_cue\}"
        \par
        7. Vary wording, word order, and synonyms across cues; do not repeat the example
        \par
        8. Do NOT mention technical terms such as "requirements", "attribute", or "metadata"
        \par\smallskip
        Output ONLY the cues, one per line, numbered 1 to \{N\}.
        &
        \raggedright\ttfamily
        You are generating diverse natural language retrieval instructions for an audio retrieval system.
        Given a spoken query, the user wants to find another clip satisfying certain requirements.
        \par\smallskip
        MATCHING constraints (same as spoken query):
        \par
        - the same \{key\} (e.g., \{values\}) as the given spoken query
        \par\smallskip
        TARGET constraints (find a specific value):
        \par
        - a specific \{key\} (e.g., \{values\}) of "\{value\}"
        \par\smallskip
        Generate \{N\} diverse and natural retrieval instructions.
        Each instruction should:
        \par
        1. Clearly indicate which attributes should MATCH the spoken query (e.g., "same emotion", "matching gender", "from the same speaker")
        \par
        2. Clearly indicate which attributes should have a SPECIFIC TARGET value (e.g., "with a happy emotion", "in a British accent")
        \par
        3. Use varied sentence structures, synonyms, and phrasings
        \par
        4. Be natural and conversational
        \par
        5. Not mention technical terms such as "requirements", "attribute", or "metadata"
        \par
        6. Refer to "the spoken query" or "the given audio" when referring to the query
        \par\smallskip
        Output ONLY the instructions, one per line, numbered 1 to \{N\}.
        \tabularnewline
        \bottomrule
    \end{tabular}
    \caption{
        Prompt templates for generating short instructions and full-sentence instructions.
        Only terminology is adapted to match the paper.
        The templates distinguish query-matching requirements in $r_q$ from specified values in $r_s$.
        Braces mark fields filled for each set of requirements, and \texttt{N} denotes the requested number of instructions.
    }
    \label{tab:instruction_generation_prompts}
\end{table*}

\begin{table*}[t]
    \centering
    \small
    \setlength{\tabcolsep}{4pt}
    \begin{tabular}{p{0.28\linewidth}p{0.20\linewidth}p{0.46\linewidth}}
        \toprule
        \textbf{Requirements} & \textbf{Short instruction} & \textbf{Full-sentence instruction} \\
        \midrule
        Same speaker
        & identical voice
        & Fetch another clip of the speaker from this audio clip. \\
        \addlinespace[2pt]
        Same environmental sound
        & matching ambiance
        & Find audio that matches the background sound environment of this recording. \\
        \addlinespace[2pt]
        Same speaker, angry speaking style
        & anger, same voice
        & Search for an angry version of this person's voice. \\
        \bottomrule
    \end{tabular}
    \caption{
        Examples of short instructions and full-sentence instructions for the same retrieval requirements.
        ``Same'' indicates a query-matching requirement in $r_q$.
        The final row also includes a specified speaking style of ``angry'' in $r_s$.
    }
    \label{tab:instruction_examples}
\end{table*}

We use Gemini~3.1~Pro Preview\footnote{\url{https://ai.google.dev/gemini-api/docs/models/gemini-3.1-pro-preview}} and Claude~Opus~4.6\footnote{\url{https://www.anthropic.com/news/claude-opus-4-6}} to express each set of requirements $r$ as natural-language instructions $z$.
Two prompt variants produce short instructions and full-sentence instructions.
We generate approximately 200 instructions per set of requirements before filtering and deduplication.

\paragraph{Prompt Variants.}
Table~\ref{tab:instruction_generation_prompts} presents the prompt templates, and Table~\ref{tab:instruction_examples} provides examples of generated instructions.
Both templates distinguish query-matching requirements in $r_q$ from specified values in $r_s$.
The short-instruction template requests phrases of two to six words, while the full-sentence template requests conversational sentences.
Both encourage diverse wording and avoid technical terms such as ``criteria'', ``attribute'', and ``metadata''.

\paragraph{Filtering.}
We retain instructions that express all requirements in $r$.
Instructions must refer to the spoken query for $r_q$ and specify the values in $r_s$.
We allow synonyms and natural paraphrases of specified values, but query-matching requirements must identify the correct attribute.
For example, accent matching must refer to ``accent'' rather than ``speaking style'' or ``way of speaking.''

\paragraph{Deduplication.}
Within each pattern of requirements, we consider instructions in random order and discard any instruction with a ROUGE-L F1 score of at least 0.7 against a previously retained instruction.

\subsection{Statistics}
\label{app:statistics}
\begin{table}[hbtp]
\centering
\small
\setlength{\tabcolsep}{3pt}
\begin{tabular}{@{}p{0.62\linewidth}rr@{}}
\toprule
\textbf{Source} & \textbf{Training} & \textbf{Validation} \\
\midrule
\multicolumn{3}{@{}l}{\textit{Real data}} \\
VoxCeleb1~\citep{voxceleb} & 294,303 & 2,974 \\
ESD~\citep{esd} & 115,282 & 1,182 \\
MELD~\citep{meld} & 72,526 & 791 \\
CREMA-D~\citep{cremad} & 56,754 & 584 \\
L2-ARCTIC~\citep{l2arctic} & 53,022 & 537 \\
LibriSpeech-C~\citep{librispeechc} & 51,553 & 521 \\
EmoV-DB~\citep{emovdb} & 49,834 & 525 \\
MSP-IMPROV~\citep{mspimprov} & 44,763 & 459 \\
VocalSound~\citep{vocalsound} & 20,678 & 209 \\
SpokenWOZ~\citep{spokenwoz} & 13,361 & 135 \\
MultiDialog~\citep{multidialog} & 12,941 & 131 \\
UrbanSound8K~\citep{urbansound8k} & 5,844 & 59 \\
EMNS~\citep{emns} & 1,022 & 10 \\
\midrule
Real data subtotal & 791,883 & 8,117 \\
\midrule
\multicolumn{3}{@{}l}{\textit{Synthetic data}} \\
MS MARCO~\citep{msmarco} & 98,395 & 1,536 \\
\midrule
\textbf{Total} & \textbf{890,278} & \textbf{9,653} \\
\bottomrule
\end{tabular}
\caption{Training and validation sample counts by source.
Each sample comprises a spoken query, an instruction, and positive and negative spoken documents.
}
\label{tab:data_counts}
\end{table}

The dataset contains 899,931 samples, comprising 800,000 from real data and 99,931 from synthetic data.
Table~\ref{tab:data_counts} reports the training and validation counts by source.

\paragraph{Data Split.}
We shuffle the samples and keep those sharing the same spoken query in a single split.
The resulting training and validation sets contain 890,278 and 9,653 samples, respectively.

\paragraph{Preventing Data Contamination.}
Our retrieval training data exclude test recordings from INSPIRE~\citep{inspire} and the 15 evaluated MAEB~\citep{maeb} tasks.
We exclude MAEB tasks derived from CREMA-D and VoxCeleb because these corpora are included in our training data.
Synthetic training uses MS MARCO text and UrbanSound8K sounds, whereas INSPIRE's Synthetic subset uses Natural Questions text and ESC-50 sounds.
The synthesis voices, text-to-speech models, and instruction-generation models also differ from those used in INSPIRE.

\clearpage
\section{Training Details}
\label{app:training}

\subsection{Setup}

\begin{table}[htbp]
    \centering
    \small
    \setlength{\tabcolsep}{4pt}
    \begin{tabular}{p{0.42\linewidth}p{0.50\linewidth}}
        \toprule
        \textbf{Setting} & \textbf{Value} \\
        \midrule
        Learning rate & $2\times10^{-4}$ \\
        Adam coefficients & $\beta_1=0.9$, $\beta_2=0.95$ \\
        Adam epsilon & $10^{-8}$ \\
        Weight decay & $0.1$ \\
        Maximum gradient norm & $1.0$ \\
        Schedule & Linear warmup, then cosine decay to zero \\
        Warmup & First 5\% of optimizer steps \\
        Training duration & 2 epochs; 3,476 optimizer steps \\
        Global batch size & 512 samples \\
        Batch size per GPU & 32 samples \\
        Contrastive temperature & $0.03$ \\
        \bottomrule
    \end{tabular}
    \caption{
        Training settings shared by both models.
        The global batch contains 32 samples on each of 16 GPUs.
    }
    \label{tab:training_settings}
\end{table}

\begin{table*}[t]
    \centering
    \footnotesize
    \setlength{\tabcolsep}{6pt}
    \begin{tabular}{p{0.47\linewidth}p{0.47\linewidth}}
        \toprule
        \textbf{Qwen2.5-Omni-3B Query Prompt:} & \textbf{Qwen2.5-Omni-3B Document Prompt:} \\
        \midrule
        \raggedright\ttfamily <|im\_start|>user\newline
        \textcolor{red}{Instruction:} \textcolor{blue}{[instruction]}\newline
        \textcolor{red}{Query:} <|audio\_bos|>\allowbreak\textcolor{blue}{[audio\_tokens]}\allowbreak<|audio\_eos|>\allowbreak<|im\_end|>\newline
        <|im\_start|>assistant
        &
        \raggedright\ttfamily <|im\_start|>user\newline
        \textcolor{red}{Document:} <|audio\_bos|>\allowbreak\textcolor{blue}{[audio\_tokens]}\allowbreak<|audio\_eos|>\allowbreak<|im\_end|>\newline
        <|im\_start|>assistant \tabularnewline
        \midrule
        \textbf{Gemma-4-E2B-it Query Prompt:} & \textbf{Gemma-4-E2B-it Document Prompt:} \\
        \midrule
        \raggedright\ttfamily <bos><|turn>user\newline
        \textcolor{red}{Instruction:} \textcolor{blue}{[instruction]}\newline
        \textcolor{red}{Query:} \textcolor{blue}{[audio\_tokens]}\allowbreak<turn|>\newline
        <|turn>model
        &
        \raggedright\ttfamily <bos><|turn>user\newline
        \textcolor{red}{Document:} \textcolor{blue}{[audio\_tokens]}\allowbreak<turn|>\newline
        <|turn>model \tabularnewline
        \midrule
        \textbf{MiniLM Query Prompt:} & \textbf{MiniLM Document Prompt:} \\
        \midrule
        \raggedright\ttfamily \textcolor{red}{Instruction:} \textcolor{blue}{[instruction]}\newline
        \textcolor{red}{Query:}
        &
        \raggedright\ttfamily \textcolor{red}{Document:} \tabularnewline
        \bottomrule
    \end{tabular}
    \caption{
        IRIS-Embed prompt templates.
        Red: input labels; blue: placeholders.
    }
    \label{tab:iris_prompts}
\end{table*}

We train IRIS-Embed using Qwen2.5-Omni-3B~\citep{qwen25omni}\footnote{\url{https://huggingface.co/Qwen/Qwen2.5-Omni-3B}} and Gemma-4-E2B-it~\citep{gemma4}\footnote{\url{https://huggingface.co/google/gemma-4-E2B-it}} with the ms-swift package~\citep{msswift}\footnote{\url{https://github.com/modelscope/ms-swift}}.
We optimize the contrastive objective in Section~\ref{subsec:objective} using the data splits in Appendix~\ref{app:statistics}.
We evaluate the final checkpoint after two epochs.

\paragraph{Optimization.}
Both models use fused AdamW with the same learning rate for all trainable components.
Weight decay is not applied to biases or normalization parameters.
Training uses bfloat16 precision and DeepSpeed ZeRO stage 2 across 16 NVIDIA GB200 GPUs.
Table~\ref{tab:training_settings} summarizes the optimization settings.

\paragraph{Trainable Components.}
The audio layer weighting module $g$ uses the pretrained \texttt{all-MiniLM-L6-v2} sentence encoder~\citep{minilm,sbert}\footnote{\url{https://huggingface.co/sentence-transformers/all-MiniLM-L6-v2}}.
Both the audio encoder and sentence encoder remain frozen.
We adapt the language model using LoRA with rank 64, scaling parameter 128, and dropout 0.05.
The audio projection and the added normalization and linear layers in $g$ are trained directly alongside the LoRA parameters.

\paragraph{LoRA Configuration.}
For Qwen2.5-Omni-3B, we apply LoRA to \texttt{q\_proj}, \texttt{k\_proj}, \texttt{v\_proj}, \texttt{o\_proj}, \texttt{gate\_proj}, \texttt{up\_proj}, and \texttt{down\_proj} across all 36 language-model layers.
For Gemma-4-E2B-it, we adapt the applicable attention and feed-forward projections across all 35 language-model layers, along with \texttt{per\_layer\_input\_gate}, \texttt{per\_layer\_projection}, and the shared \texttt{per\_layer\_model\_projection}.
Since later layers reuse keys and values, LoRA adapters for \texttt{k\_proj} and \texttt{v\_proj} are applied only to the first 15 layers.

\paragraph{Data Ordering.}
Before training, we arrange samples by source and spoken query in groups of approximately 64.
Grouping by source emphasizes within-source negatives, particularly for smaller datasets.
For dialogue continuation, this ordering is intended to reduce shortcuts based on speaker identity or recording conditions.
Grouping by spoken query places different instructions for the same query close together.
Different instructions may select different positive documents, so this ordering is intended to encourage retrieval to depend on the instruction's requirements.

\subsection{Audio Layer Weighting}

\paragraph{Computation.}
To compute $g(p)$, we mean-pool MiniLM's non-padding token hidden states, then apply learned layer normalization and a linear layer.
We apply softmax to obtain weights for the pre-encoder hidden states and all audio-encoder layer hidden states: 33 hidden-state sequences in Qwen2.5-Omni-3B and 13 in Gemma-4-E2B-it.
Each model pools and projects the weighted audio hidden states using its existing operations, including normalization, before passing them to the language model.

\paragraph{Initialization.}
We initialize the linear weights from a Gaussian distribution with mean 0 and standard deviation $10^{-4}$.
For Qwen2.5-Omni-3B, all biases are initialized to zero, yielding approximately uniform layer weights.
For Gemma-4-E2B-it, we initialize the layer weights to emphasize the final audio layer, reducing the initial deviation from its original representation.
Let $N$ denote the number of audio hidden-state sequences, including the pre-encoder hidden states.
We set the final-layer bias to
\begin{equation}
    b_{\mathrm{final}} = \log\frac{0.9(N-1)}{0.1},
\end{equation}
while initializing all other biases to zero.
For Gemma-4-E2B-it, $N=13$, so this initialization assigns approximately 90\% of the initial weight to the final layer and distributes the remainder across other layers.

\subsection{Prompts}
\label{app:prompt_templates}

Table~\ref{tab:iris_prompts} presents our prompt templates, adapted from Qwen3-VL-Embedding~\citep{qwen3vlembedding} to each model's chat and audio formats.
The query prompt includes the instruction and spoken query, while the document prompt contains the spoken document.
We extract the final-layer hidden state of the last non-padding input token as the embedding without generating a response.

For audio layer weighting, MiniLM processes text-only prompts.
The query prompt includes the instruction, while the document prompt remains fixed across all spoken documents.

\clearpage
\section{Evaluation Details}
\label{app:evaluation}

\subsection{INSPIRE}
\label{app:inspire}

\paragraph{Benchmark.}
INSPIRE~\citep{inspire}\footnote{\url{https://huggingface.co/datasets/lca0503/INSPIRE}} evaluates spoken-document retrieval given a spoken query and a natural-language instruction.
Instructions describe requirements for speech attributes.
Thus, the same spoken query may have different relevant documents depending on the instruction.
A document is relevant only if it satisfies all requirements.

\paragraph{Tasks.}
DailyTalk~\citep{dailytalk} uses the first half of a dialogue as the spoken query and evaluates retrieval of its continuation.
VCTK~\citep{vctk} evaluates speaker matching using hard negatives with the same spoken content but different speakers.
Expresso~\citep{expresso} evaluates speaker and speaking-style matching, separately or jointly, including instructions with specified speaking styles.

The Synthetic subset combines spoken queries and documents synthesized from Natural Questions~\citep{nq} using GPT-4o mini TTS~\citep{gpt4o} with environmental sounds from ESC-50~\citep{esc50}.
Synthetic evaluates answer retrieval and matching by speaker, speaking style, and environmental sound, including instructions with multiple attributes.
Speaking style and environmental sound may match the spoken query or take values specified in the instruction.
Appendix~\ref{app:statistics} describes the separation between training and evaluation data.

\paragraph{Metric.}
For each query--instruction pair, embedding methods rank documents in the corresponding subset by cosine similarity.
Recall@$K$ measures the fraction of relevant documents retrieved among the top $K$ results.
Let $Q_Z$ denote the evaluated query--instruction pairs, $\pi_K(q,z)$ the top $K$ retrieved documents, and $D^+(q,z)$ the relevant documents.
We compute average recall as
\begin{equation}
    \begin{aligned}
        &\operatorname{Recall@}K \\
        &\quad=\frac{1}{|Q_Z|}\sum_{(q,z)\in Q_Z}
        \frac{|\pi_K(q,z)\cap D^+(q,z)|}{|D^+(q,z)|}.
    \end{aligned}
\end{equation}
Each query--instruction pair contributes equally to the average.
We report Recall@$K$ as percentages for $K\in\{10,50,100\}$ on each subset.
For the instruction-type analysis in Section~\ref{subsec:instruction_types}, we compute Recall@50 separately for each instruction type.

\subsection{Comparison Settings}
\label{app:comparison_settings}
\begin{table*}[t]
    \centering
    \small
    \setlength{\tabcolsep}{4pt}
    \begin{tabular}{p{0.32\linewidth}p{0.07\linewidth}p{0.55\linewidth}}
        \toprule
        \textbf{Task} & \textbf{Type} & \textbf{Instruction} \\
        \midrule
        BeijingOpera & Clf & Find recordings with the same percussion instrument as the query. \\
        BirdCLEF & Clf & Find recordings with the same bird species as the query. \\
        Common\allowbreak Language\allowbreak Age\allowbreak Detection & Clf & Find recordings with the same speaker age group as the query. \\
        GTZANGenre & Clf & Find recordings with the same music genre as the query. \\
        IEMOCAPGender & Clf & Find recordings with the same gender as the query. \\
        MInDS14 & Clf & Find recordings with the same intent as the query. \\
        MridinghamTonic & Clf & Find recordings with the same tonic as the query. \\
        SIBFLEURS & Clf & Find recordings with the same topic as the query. \\
        VoxPopuli\allowbreak Language\allowbreak ID & Clf & Find recordings with the same spoken language as the query. \\
        \midrule
        Vehicle\allowbreak Sound\allowbreak Clustering & Clust & Find recordings with the same environmental sound as the query. \\
        VoxPopuli\allowbreak Gender\allowbreak Clustering & Clust & Find recordings with the same gender as the query. \\
        \midrule
        NMSQA\allowbreak Pair\allowbreak Classification & PC & Find recordings that contain the answer to the query. \\
        VoxPopuli\allowbreak Accent\allowbreak Pair\allowbreak Classification & PC & Find recordings with the same accent as the query. \\
        \midrule
        GTZAN\allowbreak Audio\allowbreak Reranking & Rrnk & Find recordings with the same music genre as the query. \\
        \midrule
        JamAlt\allowbreak Artist\allowbreak A2A\allowbreak Retrieval & Rtrvl & Find recordings with the same artist as the query. \\
        \bottomrule
    \end{tabular}
    \caption{
        Task instructions used to extract IRIS-Embed query embeddings for MAEB evaluation.
        Clf: classification; Clust: clustering; PC: pair classification; Rrnk: reranking; Rtrvl: retrieval.
    }
    \label{tab:maeb_instructions}
\end{table*}

Table~\ref{tab:main_results} includes published baseline results.

\paragraph{Random.}
The random baseline averages uniform random document rankings over 100 runs~\citep{inspire}.

\paragraph{General Speech Encoders.}
WavLM-large~\citep{wavlm}\footnote{\url{https://huggingface.co/microsoft/wavlm-large}} encodes spoken queries and documents without instructions.
Embeddings are obtained by mean-pooling its final-layer hidden states over time.

\paragraph{Specialized Speech Encoders.}
ECAPA-TDNN~\citep{ecapa_tdnn}\footnote{\url{https://huggingface.co/speechbrain/spkrec-ecapa-voxceleb}} and emotion2vec-base~\citep{emotion2vec}\footnote{\url{https://huggingface.co/emotion2vec/emotion2vec_base}} extract speaker and emotion embeddings from their final-layer representations, respectively.
Both encode spoken queries and documents without instructions.

\paragraph{Cascaded Retrieval.}
Whisper-large-v3~\citep{whisper}\footnote{\url{https://huggingface.co/openai/whisper-large-v3}} generates transcripts, and Qwen3-Omni-30B-A3B-Captioner~\citep{qwen3omni}\footnote{\url{https://huggingface.co/Qwen/Qwen3-Omni-30B-A3B-Captioner}} generates captions for spoken queries and documents.
Qwen3-Embedding-8B~\citep{qwen3embedding}\footnote{\url{https://huggingface.co/Qwen/Qwen3-Embedding-8B}} encodes the transcripts and captions, with the instruction included in the query input.

\paragraph{LALM Embeddings.}
Qwen2.5-Omni-3B~\citep{qwen25omni}\footnote{\url{https://huggingface.co/Qwen/Qwen2.5-Omni-3B}}, Gemma-4-E2B-it~\citep{gemma4}\footnote{\url{https://huggingface.co/google/gemma-4-E2B-it}}, and Qwen3-Omni-30B~\citep{qwen3omni}\footnote{\url{https://huggingface.co/Qwen/Qwen3-Omni-30B-A3B-Instruct}} follow INSPIRE's evaluation setup~\citep{inspire} without additional embedding training.
For these models, query and document inputs end with ``Summarize the above sentence and speech in one word'' and ``Summarize the above speech in one word'', respectively.
LCO-Embedding-Omni-3B~\citep{lcoembedding}\footnote{\url{https://huggingface.co/LCO-Embedding/LCO-Embedding-Omni-3B}} follows its published embedding extraction configuration.
All four methods extract embeddings from the final-layer hidden state of the last non-padding input token without generating a response.

\paragraph{IRIS-Embed.}
We evaluate the final IRIS-Embed checkpoints from Appendix~\ref{app:training} with learned audio layer weighting.
Table~\ref{tab:iris_prompts} gives the query and document prompt templates.
We extract the final-layer hidden state of the last non-padding input token as the embedding without generating a response.

\subsection{Layer Aggregation}
\label{app:aggregation}
Table~\ref{tab:layer} compares audio layer aggregation for Qwen2.5-Omni-3B.
Layer 13 and Final Layer use the 13th and final audio-encoder hidden states, respectively, followed by pooling and projection into the language model.
Mean Layer assigns equal weights to the pre-encoder hidden states and all layer hidden states.
Finetuned updates the audio encoder during retrieval training, while the other variants keep it frozen.
Static Weight learns a single set of layer weights shared across all queries and documents, independent of instructions or prompts.
Inst. Weight is our instruction-aware audio layer weighting method, which predicts weights from text prompts as described in Section~\ref{subsubsec:audio_layer_weighting}.
Query weights vary with the instruction, while document weights are computed from a fixed prompt.
Mean Layer, Static Weight, and Inst. Weight combine layers 0--32 for Qwen2.5-Omni-3B, with layer 0 denoting the pre-encoder hidden states.
Both weighting variants use the same instructions in the language-model query prompt.

Final Layer, Mean Layer, Static Weight, and Inst. Weight use the training data and language-model LoRA settings in Appendix~\ref{app:training} and train the audio projection.
For Gemma-4-E2B-it, we compare Final Layer, Mean Layer, and Inst. Weight.
Appendix~\ref{app:layer_aggregation} provides the full results for both models.
Appendix~\ref{app:audio_projection} separately examines frozen and trainable audio projections for Final Layer.

\subsection{MAEB}
\label{app:maeb}

\paragraph{Tasks.}
We evaluate 15 audio tasks from MAEB~\citep{maeb}\footnote{\url{https://github.com/embeddings-benchmark/mteb}}, following the task definitions and metrics of its audio-only beta leaderboard.\footnote{\url{https://mteb-leaderboard.hf.space/benchmark/MAEB(beta)}}
We exclude tasks derived from CREMA-D~\citep{cremad} and VoxCeleb~\citep{voxceleb}, as these corpora are included in our retrieval training data.
Table~\ref{tab:maeb} reports scores on a 0--100 scale using shortened task names.

\paragraph{Baseline Scores.}
We evaluate the Qwen2.5-Omni-3B and Gemma-4-E2B-it baselines with the same task instructions as IRIS-Embed.
Scores for the other baselines come from the MAEB beta leaderboard and are averaged uniformly across language subsets for each task.

\paragraph{Embedding Extraction.}
Table~\ref{tab:maeb_instructions} lists the full task names and corresponding instructions.
For all tasks except retrieval, we encode every recording with its task instruction using the query prompt in Table~\ref{tab:iris_prompts}.
Retrieval uses separate query and document prompts, as described below.
Each input passes through the full IRIS-Embed model, and we extract the final-layer hidden state of the last non-padding input token as its embedding.
Model parameters remain fixed throughout evaluation.

\paragraph{Classification.}
Following MAEB~\citep{maeb}, nine tasks evaluate classification of percussion instruments~\citep{beijingopera}, bird species~\citep{birdclef}, speaker age groups~\citep{commonlanguage}, music genres~\citep{gtzan}, gender~\citep{iemocap}, intent~\citep{minds14}, musical tonic~\citep{mridinghamtonic}, topic~\citep{sibfleurs}, and language~\citep{voxpopuli}.
Evaluation uses a logistic-regression probe fitted on fixed embeddings with eight samples per class.
Accuracy is averaged over ten resamples.

\paragraph{Clustering.}
VehicleSound~\citep{vehiclesound} and VoxPopuliGender~\citep{voxpopuli} evaluate clustering by sound class and gender, respectively.
Following MAEB, MiniBatchKMeans uses the number of reference classes as the number of clusters.
Performance is measured by V-measure~\citep{vmeasure} over bootstrapped subsamples.

\paragraph{Pair Classification.}
NMSQA~\citep{nmsqa} and VoxPopuliAccent~\citep{voxpopuli} evaluate the relevance of recording pairs.
Following MAEB, we report the highest average precision across the benchmark's similarity measures.

\paragraph{Reranking.}
GTZAN~\citep{gtzan} evaluates reranking of a provided set of relevant documents and hard negatives using cosine similarity.
We report mean average precision at 1,000, denoted MAP@1000.

\paragraph{Retrieval.}
JamAlt~\citep{jamalt} evaluates audio-to-audio retrieval of songs by the same artist as the query.
The query prompt includes the query recording and task instruction, while the document prompt contains the candidate recording without an instruction.
We report normalized discounted cumulative gain at 10, denoted nDCG@10.
\clearpage
\section{Full Results}
\label{app:full_results}

\subsection{Instruction Types}
\begin{table*}[t]
    \centering
    \small
    \renewcommand{\arraystretch}{0.90}
    \setlength{\tabcolsep}{2.5pt}
    \resizebox{0.85\linewidth}{!}{%
        \begin{tabular}{llrrrrrrrrrr}
            \toprule
            \multirow{2}{*}{\textbf{Method}} & \multirow{2}{*}{\textbf{Model}} & \textbf{DailyTalk} & \textbf{VCTK}
            & \multicolumn{4}{c}{\textbf{Expresso}} & \multicolumn{4}{c}{\textbf{Synthetic}} \\
            \cmidrule(lr){3-3}\cmidrule(lr){4-4}\cmidrule(lr){5-8}\cmidrule(l){9-12}
            & & \textbf{Cont.} & \textbf{Spk.} & \textbf{Spk.} & \textbf{Sty.} & \textbf{\shortstack{Spk.\\$+$Sty.}} & \textbf{\shortstack{Spk.\\$+$Sty.$^{*}$}} & \textbf{Ans.} & \textbf{Spk.} & \textbf{Sty.} & \textbf{Env.} \\
            \midrule
            RAND & Random & 1.06 & 1.84 & 1.73 & 1.46 & 1.38 & 1.42 & 0.84 & 1.09 & 1.38 & 1.04 \\
            \midrule
            ENC & WavLM-large & 24.50 & 35.00 & 11.50 & 30.75 & 27.50 & 1.00 & 0.75 & 0.75 & 8.75 & 7.50 \\
            \midrule
            \multirow{2}{*}{SENC} & ECAPA-TDNN & 24.00 & \textbf{100.00} & \textbf{71.50} & 24.25 & \textbf{70.50} & 0.75 & 1.67 & \textbf{87.75} & 13.50 & 1.75 \\
            & emotion2vec-base & 23.00 & 6.25 & 3.25 & 6.25 & 7.25 & 0.75 & 16.75 & 2.25 & 18.75 & 2.25 \\
            \midrule
            CAS & Qwen3-Embedding-8B & \textbf{81.50} & 6.25 & 4.00 & 5.50 & 6.00 & 6.75 & \textbf{98.50} & 1.75 & 1.00 & 4.50 \\
            \midrule
            \multirow{4}{*}{LALM} & Qwen2.5-Omni-3B & 53.00 & 3.13 & 2.00 & 6.00 & 3.75 & 3.25 & 64.75 & 1.00 & 1.25 & 0.50 \\
            & Gemma-4-E2B-it & 25.50 & 4.38 & 1.50 & 1.25 & 0.75 & 1.25 & 50.92 & 0.50 & 1.50 & 0.50 \\
            & Qwen3-Omni-30B & 74.50 & 3.13 & 2.00 & 5.25 & 4.50 & 2.50 & 80.75 & 1.25 & 2.00 & 2.00 \\
            & LCO-Embedding-Omni-3B & 75.50 & 2.50 & 2.25 & 16.25 & 8.25 & 7.00 & \underline{97.25} & 1.25 & 1.75 & 6.00 \\
            \midrule
            \multirow{2}{*}{IRIS} & Qwen2.5-Omni-3B & \underline{77.50} & \underline{85.00} & \underline{41.25} & \underline{42.00} & 62.00 & \textbf{34.75} & 90.50 & \underline{53.50} & \textbf{53.75} & \textbf{35.50} \\
            & Gemma-4-E2B-it & 44.50 & 74.38 & 33.00 & \textbf{48.00} & \underline{63.75} & \underline{30.50} & 80.58 & 51.25 & \underline{47.25} & \underline{8.50} \\
            \bottomrule
        \end{tabular}%
    }
    \par\vspace{8pt}
    \resizebox{0.95\linewidth}{!}{%
        \begin{tabular}{llrrrrrrrrrrr}
            \toprule
            \multirow{2}{*}{\textbf{Method}} & \multirow{2}{*}{\textbf{Model}} & \multicolumn{11}{c}{\textbf{Synthetic}} \\
            \cmidrule(l){3-13}
            & & \textbf{\shortstack{Spk.\\$+$Sty.}} & \textbf{\shortstack{Spk.\\$+$Sty.$^{*}$}} & \textbf{\shortstack{Spk.\\$+$Env.}} & \textbf{\shortstack{Spk.\\$+$Env.$^{*}$}} & \textbf{\shortstack{Sty.\\$+$Env.}} & \textbf{\shortstack{Sty.\\$+$Env.$^{*}$}} & \textbf{\shortstack{Sty.$^{*}$\\$+$Env.}} & \textbf{\shortstack{Spk.\\$+$Sty.\\$+$Env.}} & \textbf{\shortstack{Spk.\\$+$Sty.$^{*}$\\$+$Env.}} & \textbf{\shortstack{Spk.\\$+$Sty.\\$+$Env.$^{*}$}} & \textbf{\shortstack{Spk.\\$+$Sty.$^{*}$\\$+$Env.$^{*}$}} \\
            \midrule
            RAND & Random & 1.04 & 1.04 & 0.96 & 0.96 & 0.90 & 0.96 & 0.96 & 0.92 & 0.98 & 0.98 & 0.91 \\
            \midrule
            ENC & WavLM-large & 6.25 & 1.50 & 7.25 & 0.75 & 9.00 & 1.00 & 3.25 & 12.25 & 5.50 & 2.25 & 0.50 \\
            \midrule
            \multirow{2}{*}{SENC} & ECAPA-TDNN & \textbf{61.50} & 2.00 & 7.00 & 4.50 & 2.75 & \underline{3.75} & 0.00 & 20.50 & 1.25 & 13.50 & 1.00 \\
            & emotion2vec-base & 2.75 & 1.75 & 4.00 & 1.25 & 5.50 & 0.50 & 1.25 & 6.00 & 2.25 & 1.75 & 1.00 \\
            \midrule
            CAS & Qwen3-Embedding-8B & 2.50 & 1.50 & 3.00 & 2.75 & 2.00 & 2.75 & 3.25 & 4.25 & 1.75 & 2.25 & 2.50 \\
            \midrule
            \multirow{4}{*}{LALM} & Qwen2.5-Omni-3B & 1.50 & 0.50 & 0.50 & 0.75 & 2.00 & 1.25 & 0.25 & 0.50 & 0.75 & 1.50 & 1.00 \\
            & Gemma-4-E2B-it & 1.75 & 1.50 & 1.00 & 2.00 & 1.75 & 1.50 & 1.50 & 0.75 & 1.25 & 0.75 & 1.25 \\
            & Qwen3-Omni-30B & 1.00 & 0.50 & 0.50 & 0.75 & 1.25 & 0.75 & 0.75 & 1.00 & 1.25 & 0.75 & 1.25 \\
            & LCO-Embedding-Omni-3B & 1.75 & 1.00 & 2.75 & 3.00 & 6.25 & 3.00 & 2.50 & 4.50 & 2.75 & 4.25 & 2.50 \\
            \midrule
            \multirow{2}{*}{IRIS} & Qwen2.5-Omni-3B & \underline{54.25} & \textbf{46.00} & \textbf{47.50} & \textbf{17.25} & \textbf{35.75} & \textbf{15.00} & \textbf{38.00} & \textbf{61.75} & \textbf{57.00} & \textbf{30.00} & \textbf{26.75} \\
            & Gemma-4-E2B-it & 47.00 & \underline{45.25} & \underline{15.50} & \underline{7.25} & \underline{13.50} & 1.25 & \underline{8.25} & \underline{31.50} & \underline{18.25} & \underline{15.00} & \underline{13.75} \\
            \bottomrule
        \end{tabular}%
    }
    \caption{
        Full Recall@50 results by instruction type on INSPIRE.
        Bold and underline indicate the best and second-best results, respectively.
        Cont.: dialogue continuation; Ans.: answer retrieval; Spk.: speaker; Sty.: speaking style; Env.: environmental sound.
        $+$ combines requirements; $*$ marks a specified value.
    }
    \vspace{-8pt}
    \label{tab:instruction_results}
\end{table*}

Table~\ref{tab:instruction_results} reports Recall@50 by instruction type on INSPIRE, covering instructions with a single attribute or multiple attributes.
Relative to the corresponding LALM embedding baselines, IRIS-Embed improves Recall@50 in all 21 evaluation settings for Qwen2.5-Omni-3B and 20 of 21 for Gemma-4-E2B-it.
The exception is matching speaking style with a specified environmental sound on Synthetic, where Gemma-4-E2B-it achieves 1.25\% with IRIS-Embed versus 1.50\% with the LALM baseline.

IRIS-Embed also outperforms other baselines across several instruction types.
Both models outperform all LALM embedding baselines on speaker matching in VCTK and all four instruction types in Expresso.
Both models also outperform all comparison baselines on speaking-style matching in Expresso and Synthetic, and environmental-sound matching in Synthetic.

For instructions with multiple attributes, the Qwen2.5-Omni-3B and Gemma-4-E2B-it versions of IRIS-Embed outperform all comparison baselines in 11 and 10 of the 13 evaluation settings across Expresso and Synthetic, respectively.
On Synthetic, matching the spoken query's speaker, speaking style, and environmental sound together gives Recall@50 of 61.75\% and 31.50\%, compared with 20.50\% for the strongest baseline, ECAPA-TDNN.
Matching the speaker and a specified speaking style yields 46.00\% and 45.25\%, compared with 2.00\% for the strongest baseline.
The gains therefore extend to requirements with query-matched and explicitly specified attribute values.

\subsection{Instruction Awareness}
\label{app:instruction_awareness}
\begin{figure*}[t]
    \centering
    \includegraphics[width=\linewidth]{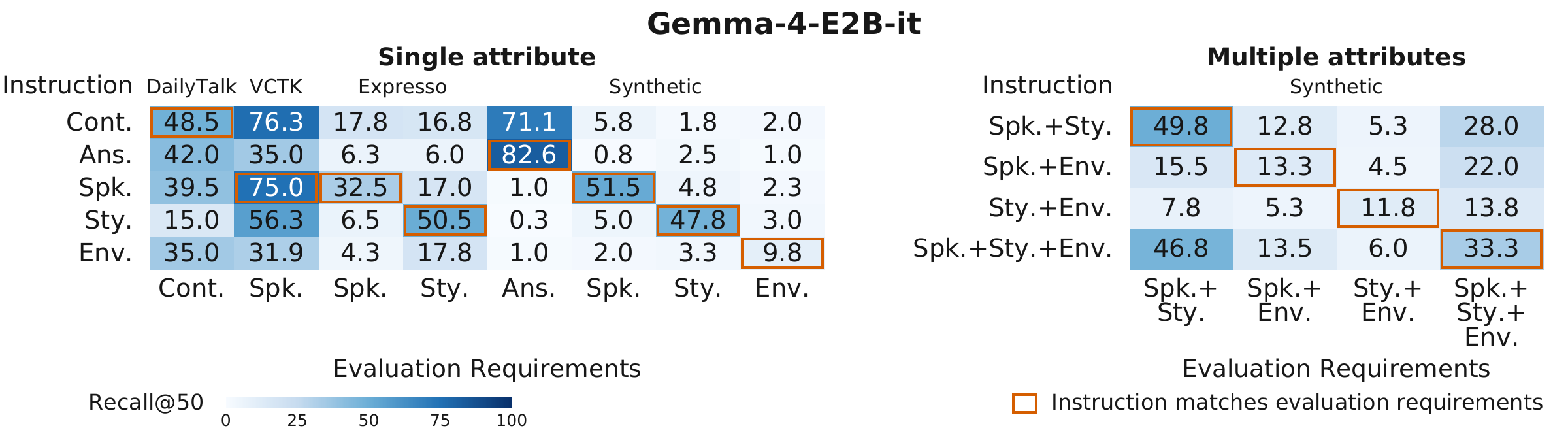}
    \caption{
        Recall@50 for IRIS-Embed with Gemma-4-E2B-it.
        Cont./Ans.: dialogue continuation/answer retrieval; Spk./Sty./Env.: speaker/speaking style/environmental sound; $+$: multiple attributes.
    }
    \label{fig:instruction_awareness_gemma}
\end{figure*}

\begin{table*}[t]
    \centering
    \small
    \setlength{\tabcolsep}{4pt}
    \begin{tabular}{lp{0.27\linewidth}p{0.60\linewidth}}
        \toprule
        \textbf{Label} & \textbf{Short instruction} & \textbf{Full-sentence instruction} \\
        \midrule
        Cont. & Continuation. & Find speech that continues the conversation in the query. \\
        \addlinespace[2pt]
        Ans. & Find the answer. & Find speech that contains the answer to the question in the query. \\
        \addlinespace[2pt]
        Spk. & Same speaker. & Find speech from the same speaker as the query. \\
        \addlinespace[2pt]
        Sty. & Same speaking style. & Find speech with the same speaking style as the query. \\
        \addlinespace[2pt]
        Env. & Same environmental sound. & Find speech with the same environmental sound as the query. \\
        \midrule
        Doc. & \multicolumn{2}{l}{\texttt{Document:}} \\
        \bottomrule
    \end{tabular}
    \caption{
        Instructions for instruction awareness and layer-weight analysis.
        The main-paper instruction-awareness and layer-weight analyses use short instructions.
        Doc. denotes the document prompt used in the layer-weight analysis.
    }
    \vspace{-8pt}
    \label{tab:awareness_instructions}
\end{table*}

We assess whether IRIS-Embed follows retrieval instructions by varying the instruction while keeping the spoken queries, document collection, and relevance labels fixed for each set of evaluation requirements.
The tables report Recall@50, with instructions in rows and evaluation requirements in columns.
Tables~\ref{tab:main_results} and~\ref{tab:instruction_results} use the original INSPIRE instructions.
Here, we replace them with fixed short or full-sentence instructions, so recall may differ from those results.

We consider five instructions with a single attribute: dialogue continuation, answer retrieval, speaker matching, speaking-style matching, and environmental-sound matching.
Dialogue continuation and answer retrieval specify different relevance requirements for spoken content.
Table~\ref{tab:awareness_instructions} lists a short instruction and a full-sentence instruction for each of the five instruction types.
The left sides of Figures~\ref{fig:instruction_awareness} and~\ref{fig:instruction_awareness_gemma} show the Short rows from Table~\ref{tab:awareness_single} for Qwen2.5-Omni-3B and Gemma-4-E2B-it, respectively.
For both models, the instruction matching the evaluation requirements achieves the highest Recall@50 across all four Synthetic evaluation settings.

\begin{table*}[t]
    \centering
    \small
    \setlength{\tabcolsep}{2pt}
    \resizebox{0.95\linewidth}{!}{%
        \begin{tabular}{p{0.38\linewidth}lrrrrrrrr}
            \toprule
            \multicolumn{10}{c}{\textbf{Qwen2.5-Omni-3B}} \\
            \midrule
            \multirow{2}{*}{\textbf{Instruction}} & \multirow{2}{*}{\textbf{Form}}
            & \textbf{DailyTalk} & \textbf{VCTK} & \multicolumn{2}{c}{\textbf{Expresso}} & \multicolumn{4}{c}{\textbf{Synthetic}} \\
            \cmidrule(lr){3-3}\cmidrule(lr){4-4}\cmidrule(lr){5-6}\cmidrule(l){7-10}
            &  & \textbf{Cont.} & \textbf{Spk.} & \textbf{Spk.} & \textbf{Sty.} & \textbf{Ans.} & \textbf{Spk.} & \textbf{Sty.} & \textbf{Env.} \\
            \midrule
            \raggedright Continuation. & Short & 79.00 & 81.88 & 19.75 & 22.50 & 87.25 & 2.75 & 2.75 & 2.75 \\
            \raggedright Find speech that continues the conversation in the query. & Full & 77.50 & 83.75 & 31.00 & 15.75 & 63.00 & 18.75 & 2.50 & 2.00 \\
            \addlinespace[2pt]
            \raggedright Find the answer. & Short & 82.50 & 73.75 & 8.00 & 4.25 & 90.25 & 2.00 & 1.75 & 1.25 \\
            \raggedright Find speech that contains the answer to the question in the query. & Full & 78.50 & 67.50 & 6.25 & 3.50 & 91.00 & 2.00 & 1.50 & 0.25 \\
            \addlinespace[2pt]
            \raggedright Same speaker. & Short & 63.50 & 85.63 & 40.75 & 13.50 & 1.50 & 53.00 & 5.50 & 0.75 \\
            \raggedright Find speech from the same speaker as the query. & Full & 67.50 & 85.63 & 41.00 & 13.50 & 1.50 & 53.00 & 6.00 & 0.75 \\
            \addlinespace[2pt]
            \raggedright Same speaking style. & Short & 10.50 & 30.00 & 4.00 & 42.50 & 0.50 & 2.75 & 55.00 & 0.50 \\
            \raggedright Find speech with the same speaking style as the query. & Full & 21.00 & 30.00 & 3.00 & 41.00 & 0.50 & 2.75 & 55.75 & 1.00 \\
            \addlinespace[2pt]
            \raggedright Same environmental sound. & Short & 20.50 & 34.38 & 4.50 & 8.75 & 1.50 & 0.75 & 0.50 & 36.75 \\
            \raggedright Find speech with the same environmental sound as the query. & Full & 29.00 & 35.63 & 4.75 & 10.00 & 1.50 & 1.00 & 0.50 & 37.25 \\
            \midrule
            \multicolumn{2}{l}{No instruction} & 75.00 & 80.63 & 20.75 & 26.00 & 74.58 & 3.25 & 3.75 & 8.75 \\
            \bottomrule
        \end{tabular}
    }
    \par\vspace{8pt}
    \resizebox{0.95\linewidth}{!}{%
        \begin{tabular}{p{0.38\linewidth}lrrrrrrrr}
            \toprule
            \multicolumn{10}{c}{\textbf{Gemma-4-E2B-it}} \\
            \midrule
            \multirow{2}{*}{\textbf{Instruction}} & \multirow{2}{*}{\textbf{Form}}
            & \textbf{DailyTalk} & \textbf{VCTK} & \multicolumn{2}{c}{\textbf{Expresso}} & \multicolumn{4}{c}{\textbf{Synthetic}} \\
            \cmidrule(lr){3-3}\cmidrule(lr){4-4}\cmidrule(lr){5-6}\cmidrule(l){7-10}
            &  & \textbf{Cont.} & \textbf{Spk.} & \textbf{Spk.} & \textbf{Sty.} & \textbf{Ans.} & \textbf{Spk.} & \textbf{Sty.} & \textbf{Env.} \\
            \midrule
            \raggedright Continuation. & Short & 48.50 & 76.25 & 17.75 & 16.75 & 71.08 & 5.75 & 1.75 & 2.00 \\
            \raggedright Find speech that continues the conversation in the query. & Full & 49.00 & 75.63 & 19.25 & 13.75 & 73.83 & 3.50 & 1.25 & 1.25 \\
            \addlinespace[2pt]
            \raggedright Find the answer. & Short & 42.00 & 35.00 & 6.25 & 6.00 & 82.58 & 0.75 & 2.50 & 1.00 \\
            \raggedright Find speech that contains the answer to the question in the query. & Full & 41.50 & 36.25 & 6.75 & 6.00 & 83.08 & 1.00 & 2.25 & 1.25 \\
            \addlinespace[2pt]
            \raggedright Same speaker. & Short & 39.50 & 75.00 & 32.50 & 17.00 & 1.00 & 51.50 & 4.75 & 2.25 \\
            \raggedright Find speech from the same speaker as the query. & Full & 45.50 & 75.63 & 32.25 & 16.75 & 1.50 & 52.75 & 4.75 & 2.50 \\
            \addlinespace[2pt]
            \raggedright Same speaking style. & Short & 15.00 & 56.25 & 6.50 & 50.50 & 0.25 & 5.00 & 47.75 & 3.00 \\
            \raggedright Find speech with the same speaking style as the query. & Full & 22.50 & 54.38 & 7.25 & 47.25 & 0.25 & 4.50 & 46.75 & 2.50 \\
            \addlinespace[2pt]
            \raggedright Same environmental sound. & Short & 35.00 & 31.88 & 4.25 & 17.75 & 1.00 & 2.00 & 3.25 & 9.75 \\
            \raggedright Find speech with the same environmental sound as the query. & Full & 41.00 & 35.63 & 5.00 & 14.00 & 1.00 & 1.75 & 2.50 & 9.50 \\
            \midrule
            \multicolumn{2}{l}{No instruction} & 44.00 & 75.63 & 20.00 & 39.00 & 8.75 & 21.25 & 24.25 & 3.75 \\
            \bottomrule
        \end{tabular}
    }
    \caption{
        Recall@50 for IRIS-Embed with short instructions, full-sentence instructions, or no instruction.
        Rows show the exact instructions; Short/Full denote short/full-sentence instructions.
        Cont./Ans.: dialogue continuation/answer retrieval; Spk./Sty./Env.: speaker/speaking style/environmental sound.
    }
    \label{tab:awareness_single}
\end{table*}

Table~\ref{tab:awareness_single} reports both wording forms, labeled Short and Full, and a no-instruction condition.
Within each wording form, the matching instruction achieves the highest Recall@50 for both models across all four Synthetic evaluation settings and for speaker and speaking-style matching on Expresso.
The matching instruction also performs best for Qwen2.5-Omni-3B on VCTK and Gemma-4-E2B-it on DailyTalk.
Both short instructions and full-sentence instructions therefore guide retrieval according to the requested requirements in most evaluation settings.

The matching instruction does not always achieve the highest recall.
On DailyTalk, Qwen2.5-Omni-3B performs best with the answer-retrieval instruction under both wording forms.
This may reflect overlap between answering a question and continuing a dialogue.
On VCTK, Gemma-4-E2B-it achieves 76.25\% with the short continuation instruction and 75.00\% with the short speaker-matching instruction.
The corresponding full-sentence instructions and no instruction all yield 75.63\%, suggesting that speaker information remains available without an explicit speaker-matching instruction.

\begin{table*}[t]
    \centering
    \small
    \setlength{\tabcolsep}{2pt}
    \resizebox{0.95\linewidth}{!}{%
        \begin{tabular}{p{0.40\linewidth}rrrrrrrr}
            \toprule
            \multirow{2}{*}{\textbf{Instruction}} & \multicolumn{4}{c}{\textbf{Qwen2.5-Omni-3B}} & \multicolumn{4}{c}{\textbf{Gemma-4-E2B-it}} \\
            \cmidrule(lr){2-5}\cmidrule(l){6-9}
            & \textbf{\shortstack{Spk.\\$+$Sty.}} & \textbf{\shortstack{Spk.\\$+$Env.}} & \textbf{\shortstack{Sty.\\$+$Env.}} & \textbf{\shortstack{Spk.\\$+$Sty.\\$+$Env.}} & \textbf{\shortstack{Spk.\\$+$Sty.}} & \textbf{\shortstack{Spk.\\$+$Env.}} & \textbf{\shortstack{Sty.\\$+$Env.}} & \textbf{\shortstack{Spk.\\$+$Sty.\\$+$Env.}} \\
            \midrule
            Find the answer. & 0.75 & 1.75 & 3.25 & 2.00 & 1.00 & 1.25 & 1.75 & 1.50 \\
            \addlinespace[2pt]
            Continuation. & 1.50 & 4.50 & 5.25 & 6.50 & 2.75 & 3.75 & 3.25 & 4.50 \\
            \addlinespace[2pt]
            Same speaker. & 18.50 & 8.50 & 2.50 & 10.25 & 18.00 & 8.50 & 4.00 & 14.25 \\
            \addlinespace[2pt]
            Same speaking style. & 9.50 & 2.25 & 3.25 & 7.00 & 12.25 & 4.50 & 7.75 & 12.25 \\
            \addlinespace[2pt]
            Same environmental sound. & 1.50 & 10.75 & 15.00 & 11.75 & 2.50 & 6.75 & 9.25 & 10.50 \\
            \midrule
            Same speaker and same speaking style. & 52.00 & 11.50 & 7.50 & 27.25 & 49.75 & 12.75 & 5.25 & 28.00 \\
            \addlinespace[2pt]
            Same speaker and same environmental sound. & 8.25 & 45.25 & 11.75 & 45.25 & 15.50 & 13.25 & 4.50 & 22.00 \\
            \addlinespace[2pt]
            Same speaking style and same environmental sound. & 6.00 & 11.50 & 35.00 & 36.75 & 7.75 & 5.25 & 11.75 & 13.75 \\
            \addlinespace[2pt]
            Same speaker and same speaking style and same environmental sound. & 30.00 & 29.25 & 22.25 & 65.00 & 46.75 & 13.50 & 6.00 & 33.25 \\
            \midrule
            No instruction & 6.00 & 13.25 & 9.50 & 17.75 & 35.75 & 11.25 & 7.75 & 30.25 \\
            \bottomrule
        \end{tabular}
    }
    \caption{
        Recall@50 for IRIS-Embed under short instructions and no instruction.
        Columns evaluate Synthetic requirements involving multiple attributes.
        Rows show the exact instructions; No instruction indicates an empty instruction.
        Spk., Sty., and Env. match the spoken query's speaker, speaking style, and environmental sound; $+$ combines requirements.
    }
    \label{tab:awareness_combined}
\end{table*}

We further evaluate four short instructions with multiple attributes of the spoken query.
These cover all pairwise combinations of speaker, speaking style, and environmental sound, as well as all three attributes together.
The right panels of Figures~\ref{fig:instruction_awareness} and~\ref{fig:instruction_awareness_gemma} show the rows for these four short instructions in Table~\ref{tab:awareness_combined} across the corresponding Synthetic evaluation settings.
The matching instruction achieves the highest Recall@50 in all four evaluation settings for Qwen2.5-Omni-3B and three of four for Gemma-4-E2B-it.
When relevance requires matching all three attributes, the corresponding instruction outperforms all pairwise instructions for both models.

Table~\ref{tab:awareness_combined} provides the full comparison, including the five short instructions above and no instruction.
For both models, the matching instruction outperforms these additional conditions in all four evaluation settings.
These findings show that IRIS-Embed responds to instructions with multiple attributes.

\subsection{Layer Aggregation}
\label{app:layer_aggregation}
\begin{table*}[t]
    \centering
    \small
    \setlength{\tabcolsep}{2.8pt}

    \resizebox{\linewidth}{!}{%
        \begin{tabular}{l|rrrrrrrrrrrr|rrr}
            \toprule

            \multirow{2}{*}{\textbf{Layer Aggregation}}
            & \multicolumn{3}{c}{\textbf{DailyTalk}}
            & \multicolumn{3}{c}{\textbf{VCTK}}
            & \multicolumn{3}{c}{\textbf{Expresso}}
            & \multicolumn{3}{c|}{\textbf{Synthetic}}
            & \multicolumn{3}{c}{\textbf{Avg.}}
            \\
            \cmidrule(lr){2-4}\cmidrule(lr){5-7}\cmidrule(lr){8-10}\cmidrule(lr){11-13}\cmidrule(l){14-16}
            &
            \textbf{R@10} & \textbf{R@50} & \textbf{R@100}
            & \textbf{R@10} & \textbf{R@50} & \textbf{R@100}
            & \textbf{R@10} & \textbf{R@50} & \textbf{R@100}
            & \textbf{R@10} & \textbf{R@50} & \textbf{R@100}
            & \textbf{R@10} & \textbf{R@50} & \textbf{R@100}
            \\

            \midrule

            \multicolumn{16}{l}{Qwen2.5-Omni-3B} \\

            \quad Layer 13
            & 5.50 & 15.50 & 22.50
            & 61.88 & 76.88 & 84.38
            & \textbf{32.44} & \textbf{48.94} & \textbf{57.75}
            & 17.95 & 35.27 & 46.12
            & 29.44 & 44.15 & 52.69
            \\

            \quad Final Layer
            & 59.50 & \textbf{82.50} & 85.50
            & 34.38 & 48.75 & 59.38
            & 15.88 & 28.25 & 36.88
            & 18.47 & 34.72 & 45.20
            & 32.06 & 48.56 & 56.74
            \\

            \quad Mean Layer
            & 47.50 & 70.00 & 78.50
            & 57.50 & 71.25 & 81.25
            & 25.75 & 43.13 & 52.63
            & 26.03 & 44.90 & 55.00
            & 39.20 & 57.32 & 66.85
            \\

            \quad Finetuned
            & 14.50 & 32.50 & 42.00
            & \textbf{76.25} & \textbf{91.25} & \textbf{95.00}
            & 31.19 & 47.38 & 55.50
            & 24.90 & 43.35 & 54.49
            & 36.71 & 53.62 & 61.75
            \\

            \quad Static Weight
            & 45.00 & 66.00 & 76.50
            & 61.88 & 78.75 & 87.50
            & 25.88 & 43.19 & 55.31
            & \textbf{26.47} & \textbf{45.28} & \textbf{56.13}
            & 39.81 & 58.31 & 68.86
            \\

            \quad Inst. Weight (Ours)
            & \textbf{60.00} & 77.50 & \textbf{87.50}
            & 67.50 & 85.00 & 89.38
            & 29.63 & 45.00 & 53.88
            & 26.06 & 44.17 & 54.78
            & \textbf{45.80} & \textbf{62.92} & \textbf{71.39}
            \\

            \midrule

            \multicolumn{16}{l}{Gemma-4-E2B-it} \\

            \quad Final Layer
            & 24.50 & 42.50 & 52.00
            & 30.63 & 48.13 & 59.38
            & 17.06 & 30.00 & 37.69
            & 12.01 & 21.70 & 29.72
            & 21.05 & 35.58 & 44.70
            \\

            \quad Mean Layer
            & 12.50 & 25.00 & 37.00
            & 52.50 & 69.38 & 78.13
            & \textbf{32.00} & \textbf{46.06} & \textbf{54.63}
            & 11.63 & 21.65 & 29.38
            & 27.16 & 40.52 & 49.79
            \\

            \quad Inst. Weight (Ours)
            & \textbf{25.50} & \textbf{44.50} & \textbf{54.50}
            & \textbf{58.75} & \textbf{74.38} & \textbf{79.38}
            & 28.63 & 43.81 & 52.31
            & \textbf{15.07} & \textbf{26.94} & \textbf{36.08}
            & \textbf{31.99} & \textbf{47.41} & \textbf{55.57}
            \\

            \bottomrule
        \end{tabular}%
    }

    \caption{
        Full Recall@$K$ for audio layer aggregation on INSPIRE.
        Avg.: macro-average across the four subsets.
        Inst. Weight denotes instruction-aware audio layer weighting.
        Static Weight uses trainable weights independent of instructions and prompts.
        Bold marks column maxima within each backbone.
    }
    \label{tab:layer_full}

\end{table*}

\begin{table*}[t]
    \centering
    \small
    \setlength{\tabcolsep}{2pt}
    \resizebox{0.85\linewidth}{!}{%
        \begin{tabular}{lrrrrrrrrrr}
            \toprule
            \multirow{2}{*}{\textbf{Layer Aggregation}} & \textbf{DailyTalk} & \textbf{VCTK}
            & \multicolumn{4}{c}{\textbf{Expresso}} & \multicolumn{4}{c}{\textbf{Synthetic}} \\
            \cmidrule(lr){2-2}\cmidrule(lr){3-3}\cmidrule(lr){4-7}\cmidrule(l){8-11}
            & \textbf{Cont.} & \textbf{Spk.} & \textbf{Spk.} & \textbf{Sty.} & \textbf{\shortstack{Spk.\\$+$Sty.}} & \textbf{\shortstack{Spk.\\$+$Sty.$^{*}$}} & \textbf{Ans.} & \textbf{Spk.} & \textbf{Sty.} & \textbf{Env.} \\
            \midrule
            \multicolumn{11}{l}{Qwen2.5-Omni-3B} \\
            \quad Layer 13 & 15.50 & 76.88 & 44.75 & \textbf{45.50} & \textbf{67.25} & \textbf{38.25} & 2.00 & 43.50 & \textbf{54.75} & 39.25 \\
            \quad Final Layer & \textbf{82.50} & 48.75 & 11.25 & 39.75 & 44.75 & 17.25 & \textbf{94.50} & 27.50 & 40.50 & 31.00 \\
            \quad Mean Layer & 70.00 & 71.25 & 37.75 & 42.00 & 57.75 & 35.00 & 87.75 & 50.75 & 45.50 & 40.50 \\
            \quad Finetuned & 32.50 & \textbf{91.25} & \textbf{51.50} & 42.25 & 65.00 & 30.75 & 68.25 & 52.50 & 54.50 & 43.25 \\
            \quad Static Weight & 66.00 & 78.75 & 33.00 & 41.75 & 61.50 & 36.50 & 87.00 & 48.50 & 52.25 & \textbf{45.50} \\
            \quad Inst. Weight (Ours) & 77.50 & 85.00 & 41.25 & 42.00 & 62.00 & 34.75 & 90.50 & \textbf{53.50} & 53.75 & 35.50 \\
            \midrule
            \multicolumn{11}{l}{Gemma-4-E2B-it} \\
            \quad Final Layer & 42.50 & 48.13 & 13.50 & 45.50 & 43.25 & 17.75 & \textbf{83.25} & 28.75 & 37.50 & 7.75 \\
            \quad Mean Layer & 25.00 & 69.38 & 32.75 & 42.25 & \textbf{72.50} & \textbf{36.75} & 1.50 & \textbf{54.50} & 44.50 & 4.00 \\
            \quad Inst. Weight (Ours) & \textbf{44.50} & \textbf{74.38} & \textbf{33.00} & \textbf{48.00} & 63.75 & 30.50 & 80.58 & 51.25 & \textbf{47.25} & \textbf{8.50} \\
            \bottomrule
        \end{tabular}
    }
    \par\vspace{8pt}
    \resizebox{0.95\linewidth}{!}{%
        \begin{tabular}{lrrrrrrrrrrr}
            \toprule
            \multirow{2}{*}{\textbf{Layer Aggregation}} & \multicolumn{11}{c}{\textbf{Synthetic}} \\
            \cmidrule(l){2-12}
            & \textbf{\shortstack{Spk.\\$+$Sty.}} & \textbf{\shortstack{Spk.\\$+$Sty.$^{*}$}} & \textbf{\shortstack{Spk.\\$+$Env.}} & \textbf{\shortstack{Spk.\\$+$Env.$^{*}$}} & \textbf{\shortstack{Sty.\\$+$Env.}} & \textbf{\shortstack{Sty.\\$+$Env.$^{*}$}} & \textbf{\shortstack{Sty.$^{*}$\\$+$Env.}} & \textbf{\shortstack{Spk.\\$+$Sty.\\$+$Env.}} & \textbf{\shortstack{Spk.\\$+$Sty.$^{*}$\\$+$Env.}} & \textbf{\shortstack{Spk.\\$+$Sty.\\$+$Env.$^{*}$}} & \textbf{\shortstack{Spk.\\$+$Sty.$^{*}$\\$+$Env.$^{*}$}} \\
            \midrule
            \multicolumn{12}{l}{Qwen2.5-Omni-3B} \\
            \quad Layer 13 & 44.50 & 40.50 & 46.00 & 15.00 & 37.75 & 8.50 & 34.50 & 63.00 & 58.50 & 22.50 & 18.75 \\
            \quad Final Layer & 30.75 & 27.25 & 35.75 & 19.50 & 28.50 & 14.00 & 29.50 & 45.75 & 38.50 & 32.50 & 25.25 \\
            \quad Mean Layer & 49.75 & 43.75 & \textbf{55.00} & 20.75 & 35.75 & 16.50 & 38.75 & 65.50 & 56.25 & \textbf{38.75} & \textbf{28.25} \\
            \quad Finetuned & 49.75 & 43.50 & 51.25 & 19.50 & 35.75 & 11.00 & \textbf{39.50} & \textbf{67.50} & \textbf{62.75} & 32.50 & 18.75 \\
            \quad Static Weight & 48.25 & 41.25 & 50.25 & \textbf{22.00} & \textbf{40.50} & \textbf{18.25} & 37.25 & 67.00 & 57.25 & 37.25 & 26.75 \\
            \quad Inst. Weight (Ours) & \textbf{54.25} & \textbf{46.00} & 47.50 & 17.25 & 35.75 & 15.00 & 38.00 & 61.75 & 57.00 & 30.00 & 26.75 \\
            \midrule
            \multicolumn{12}{l}{Gemma-4-E2B-it} \\
            \quad Final Layer & 30.50 & 24.00 & 14.50 & \textbf{9.25} & 12.25 & \textbf{5.00} & \textbf{10.50} & 24.00 & 13.25 & 14.00 & 11.00 \\
            \quad Mean Layer & \textbf{55.50} & \textbf{50.00} & 13.25 & 6.00 & 4.50 & 3.75 & 3.50 & 29.25 & \textbf{25.00} & 14.00 & \textbf{15.50} \\
            \quad Inst. Weight (Ours) & 47.00 & 45.25 & \textbf{15.50} & 7.25 & \textbf{13.50} & 1.25 & 8.25 & \textbf{31.50} & 18.25 & \textbf{15.00} & 13.75 \\
            \bottomrule
        \end{tabular}
    }
    \caption{
        Recall@50 by instruction type for audio layer aggregation on INSPIRE.
        Inst. Weight denotes instruction-aware audio layer weighting.
        Static Weight uses trainable weights independent of instructions and prompts.
        Instruction labels follow Table~\ref{tab:instruction_results}; bold marks column maxima within each backbone.
    }
    \label{tab:layer_instruction_results}
\end{table*}

Table~\ref{tab:layer_full} reports Recall@10, Recall@50, and Recall@100 for audio layer aggregation, while Table~\ref{tab:layer_instruction_results} provides Recall@50 by instruction type.
For Qwen2.5-Omni-3B, the comparison includes Layer 13, Final Layer, Mean Layer, Finetuned, Static Weight, and Inst. Weight.
For Gemma-4-E2B-it, we compare Final Layer, Mean Layer, and our method.
Appendix~\ref{app:aggregation} describes these methods and their training settings.

For Qwen2.5-Omni-3B, our method achieves the highest macro-average Recall@50 of 62.92\% in Table~\ref{tab:layer}, a gain of 14.36 percentage points over Final Layer.
It consistently outperforms Final Layer on VCTK, Expresso, and Synthetic.
Final Layer remains higher on DailyTalk at Recall@50.

Layer 13 and Finetuned improve speaker and speaking-style matching over Final Layer, but have much lower dialogue-continuation recall.
Finetuned consistently leads on VCTK and Layer 13 on Expresso.
Our method consistently outperforms both variants on DailyTalk and Synthetic.
These methods trade off performance across retrieval requirements.

Our method also exceeds Mean Layer and Static Weight in Recall@50 on DailyTalk, VCTK, and Expresso.
The macro-average Recall@50 gains are 5.60 and 4.61 percentage points, respectively.
Both alternatives remain higher on Synthetic at Recall@50.
Static Weight also exceeds our method on Synthetic at Recall@10 and Recall@100, and on Expresso at Recall@100.

The instruction-type results further distinguish these methods.
On answer retrieval, our method achieves 90.50\% Recall@50, compared with 68.25\% for Finetuned and 2.00\% for Layer 13.
In the pairwise comparison with Static Weight, our method has higher Recall@50 in seven of eight evaluation settings with a single attribute.
For instructions with multiple attributes, Static Weight has higher recall in eight of thirteen settings, our method in four, and one is tied.

With Gemma-4-E2B-it, our method consistently outperforms Final Layer on VCTK, Expresso, and Synthetic, while Mean Layer leads on Expresso.
Answer retrieval Recall@50 increases from 1.50\% with Mean Layer to 80.58\% with our method.
However, Mean Layer performs better on several instructions that combine speaker matching with other attributes.
The aggregate gains support the effectiveness of instruction-aware audio layer weighting with a frozen audio encoder.

\subsection{Audio Projection}
\label{app:audio_projection}
\begin{table*}[t]
    \centering
    \small
    \setlength{\tabcolsep}{2.8pt}
    \resizebox{\linewidth}{!}{%
        \begin{tabular}{l|rrrrrrrrrrrr|rrr}
            \toprule
            \multirow{2}{*}{\textbf{Audio Projection}}
            & \multicolumn{3}{c}{\textbf{DailyTalk}}
            & \multicolumn{3}{c}{\textbf{VCTK}}
            & \multicolumn{3}{c}{\textbf{Expresso}}
            & \multicolumn{3}{c|}{\textbf{Synthetic}}
            & \multicolumn{3}{c}{\textbf{Avg.}} \\
            \cmidrule(lr){2-4}\cmidrule(lr){5-7}\cmidrule(lr){8-10}\cmidrule(lr){11-13}\cmidrule(l){14-16}
            & \textbf{R@10} & \textbf{R@50} & \textbf{R@100}
            & \textbf{R@10} & \textbf{R@50} & \textbf{R@100}
            & \textbf{R@10} & \textbf{R@50} & \textbf{R@100}
            & \textbf{R@10} & \textbf{R@50} & \textbf{R@100}
            & \textbf{R@10} & \textbf{R@50} & \textbf{R@100} \\
            \midrule
            \multicolumn{16}{l}{Qwen2.5-Omni-3B} \\
            \quad Frozen
            & \textbf{69.00} & \textbf{83.00} & \textbf{87.50}
            & 24.38 & 41.88 & 56.88
            & 14.13 & 26.50 & 36.50
            & 16.83 & 30.80 & 42.00
            & 31.09 & 45.55 & 55.72 \\
            \quad Trainable
            & 59.50 & 82.50 & 85.50
            & \textbf{34.38} & \textbf{48.75} & \textbf{59.38}
            & \textbf{15.88} & \textbf{28.25} & \textbf{36.88}
            & \textbf{18.47} & \textbf{34.72} & \textbf{45.20}
            & \textbf{32.06} & \textbf{48.56} & \textbf{56.74} \\
            \midrule
            \multicolumn{16}{l}{Gemma-4-E2B-it} \\
            \quad Frozen
            & 17.00 & 34.00 & 46.00
            & 24.38 & 45.00 & 58.13
            & 13.63 & 24.69 & 31.75
            & 9.06 & 15.83 & 21.78
            & 16.02 & 29.88 & 39.42 \\
            \quad Trainable
            & \textbf{24.50} & \textbf{42.50} & \textbf{52.00}
            & \textbf{30.63} & \textbf{48.13} & \textbf{59.38}
            & \textbf{17.06} & \textbf{30.00} & \textbf{37.69}
            & \textbf{12.01} & \textbf{21.70} & \textbf{29.72}
            & \textbf{21.05} & \textbf{35.58} & \textbf{44.70} \\
            \bottomrule
        \end{tabular}%
    }
    \caption{
        Recall@$K$ on INSPIRE using only the final audio-encoder hidden states.
        Avg.: macro-average across the four subsets.
        Frozen and Trainable indicate whether the audio projection is frozen or updated during retrieval training.
        Bold marks column maxima within each backbone.
    }
    \label{tab:projection}
\end{table*}

\begin{table*}[t]
    \centering
    \small
    \setlength{\tabcolsep}{2pt}
    \resizebox{0.85\linewidth}{!}{%
        \begin{tabular}{lrrrrrrrrrr}
            \toprule
            \multirow{2}{*}{\textbf{Audio Projection}} & \textbf{DailyTalk} & \textbf{VCTK}
            & \multicolumn{4}{c}{\textbf{Expresso}} & \multicolumn{4}{c}{\textbf{Synthetic}} \\
            \cmidrule(lr){2-2}\cmidrule(lr){3-3}\cmidrule(lr){4-7}\cmidrule(l){8-11}
            & \textbf{Cont.} & \textbf{Spk.} & \textbf{Spk.} & \textbf{Sty.} & \textbf{\shortstack{Spk.\\$+$Sty.}} & \textbf{\shortstack{Spk.\\$+$Sty.$^{*}$}} & \textbf{Ans.} & \textbf{Spk.} & \textbf{Sty.} & \textbf{Env.} \\
            \midrule
            \multicolumn{11}{l}{Qwen2.5-Omni-3B} \\
            \quad Frozen & \textbf{83.00} & 41.88 & \textbf{14.50} & 34.25 & 39.25 & \textbf{18.00} & 92.25 & 24.00 & 39.75 & 29.25 \\
            \quad Trainable & 82.50 & \textbf{48.75} & 11.25 & \textbf{39.75} & \textbf{44.75} & 17.25 & \textbf{94.50} & \textbf{27.50} & \textbf{40.50} & \textbf{31.00} \\
            \midrule
            \multicolumn{11}{l}{Gemma-4-E2B-it} \\
            \quad Frozen & 34.00 & 45.00 & 9.75 & 35.25 & 37.75 & 16.00 & \textbf{84.75} & 17.25 & \textbf{39.25} & 0.75 \\
            \quad Trainable & \textbf{42.50} & \textbf{48.13} & \textbf{13.50} & \textbf{45.50} & \textbf{43.25} & \textbf{17.75} & 83.25 & \textbf{28.75} & 37.50 & \textbf{7.75} \\
            \bottomrule
        \end{tabular}
    }
    \par\vspace{8pt}
    \resizebox{0.95\linewidth}{!}{%
        \begin{tabular}{lrrrrrrrrrrr}
            \toprule
            \multirow{2}{*}{\textbf{Audio Projection}} & \multicolumn{11}{c}{\textbf{Synthetic}} \\
            \cmidrule(l){2-12}
            & \textbf{\shortstack{Spk.\\$+$Sty.}} & \textbf{\shortstack{Spk.\\$+$Sty.$^{*}$}} & \textbf{\shortstack{Spk.\\$+$Env.}} & \textbf{\shortstack{Spk.\\$+$Env.$^{*}$}} & \textbf{\shortstack{Sty.\\$+$Env.}} & \textbf{\shortstack{Sty.\\$+$Env.$^{*}$}} & \textbf{\shortstack{Sty.$^{*}$\\$+$Env.}} & \textbf{\shortstack{Spk.\\$+$Sty.\\$+$Env.}} & \textbf{\shortstack{Spk.\\$+$Sty.$^{*}$\\$+$Env.}} & \textbf{\shortstack{Spk.\\$+$Sty.\\$+$Env.$^{*}$}} & \textbf{\shortstack{Spk.\\$+$Sty.$^{*}$\\$+$Env.$^{*}$}} \\
            \midrule
            \multicolumn{12}{l}{Qwen2.5-Omni-3B} \\
            \quad Frozen & 26.75 & 23.00 & 30.00 & 17.00 & 26.50 & 12.00 & 28.50 & 39.00 & 30.25 & 25.75 & 18.00 \\
            \quad Trainable & \textbf{30.75} & \textbf{27.25} & \textbf{35.75} & \textbf{19.50} & \textbf{28.50} & \textbf{14.00} & \textbf{29.50} & \textbf{45.75} & \textbf{38.50} & \textbf{32.50} & \textbf{25.25} \\
            \midrule
            \multicolumn{12}{l}{Gemma-4-E2B-it} \\
            \quad Frozen & 26.25 & 19.50 & 5.00 & 3.50 & 3.00 & 2.50 & 3.75 & 11.25 & 6.25 & 6.50 & 8.00 \\
            \quad Trainable & \textbf{30.50} & \textbf{24.00} & \textbf{14.50} & \textbf{9.25} & \textbf{12.25} & \textbf{5.00} & \textbf{10.50} & \textbf{24.00} & \textbf{13.25} & \textbf{14.00} & \textbf{11.00} \\
            \bottomrule
        \end{tabular}
    }
    \caption{
        Recall@50 by instruction type on INSPIRE using only the final audio-encoder hidden states.
        Frozen and Trainable indicate whether the audio projection is frozen or updated during retrieval training.
        Instruction labels follow Table~\ref{tab:instruction_results}.
        Bold marks column maxima within each backbone.
    }
    \label{tab:projection_instruction_results}
\end{table*}

We examine whether training the audio projection improves retrieval using the final audio-encoder hidden states.
Both variants freeze the audio encoder and train the language-model LoRA parameters.
Frozen keeps the audio projection fixed, while Trainable updates it during retrieval training.
Tables~\ref{tab:projection} and~\ref{tab:projection_instruction_results} report results by subset and instruction type, respectively.

For both models, the trainable audio projection consistently outperforms the frozen projection on VCTK, Expresso, and Synthetic.
On Synthetic, these gains cover all eleven instruction types with multiple attributes at Recall@50.
The effects on spoken-content retrieval vary between models.
At Recall@50, Qwen2.5-Omni-3B improves on answer retrieval but declines on dialogue continuation, while Gemma-4-E2B-it shows the opposite trend.

With a trainable audio projection in both settings, Weighted Layer consistently outperforms Final Layer on VCTK, Expresso, and Synthetic for both models.
Thus, the observed gains over Final Layer persist when both variants train the audio projection.

\subsection{Layer Weights}
\label{app:layer_weights}
\begin{figure*}[t]
    \centering
    \includegraphics[width=\linewidth]{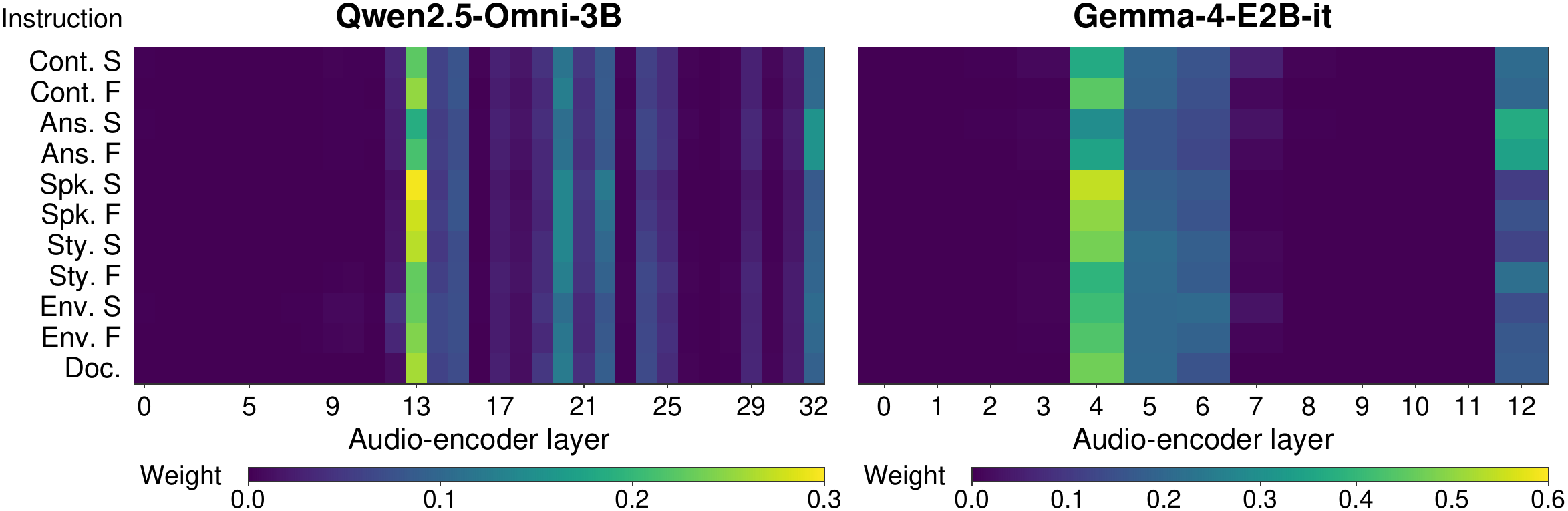}
    \caption{
        Audio layer weights adjusted for hidden-state scale.
        S/F: short/full-sentence instructions; Doc.: document prompt; 0: pre-encoder hidden states.
        Encoder layers are indexed from one.
        Color scales differ between models.
    }
    \label{fig:full_layer_weights}
\end{figure*}

We analyze how audio layer weights vary across retrieval requirements and instruction wording using the ten instructions in Table~\ref{tab:awareness_instructions}.
The document prompt serves as a reference without an instruction.
All prompts follow the MiniLM templates in Table~\ref{tab:iris_prompts}.
Figure~\ref{fig:full_layer_weights} presents all eleven prompts for both models, while Figure~\ref{fig:layer_weights} focuses on the five short instructions and the document prompt for Qwen2.5-Omni-3B.
For Gemma-4-E2B-it, layer 4 receives the highest adjusted weight under every query instruction except the short answer-retrieval instruction, which places the highest weight on the final layer.

To account for differences in hidden-state scale, we adjust the layer weights using the same 680 unique INSPIRE spoken queries for every prompt and both models.
Let $w_\ell(p)$ denote the original weight for layer $\ell$ under text prompt $p$.
For a recording $x$, let $s_\ell(x)$ denote the root mean square of the hidden states across hidden dimensions, averaged over valid frames.
We compute the adjusted weights as
\begin{equation*}
    \widehat{w}_\ell(x,p)
    = \frac{w_\ell(p)s_\ell(x)}{\sum_k w_k(p)s_k(x)},
\end{equation*}
where $k$ and $\ell$ index the pre-encoder hidden states and all audio-encoder layer hidden states.
The adjusted weights sum to one for each recording and prompt.
We first average them within each INSPIRE subset, then average the four subset means equally.
This adjustment is used only for analysis and does not affect model computation.
The adjusted weights describe the relative scales of weighted audio layer hidden states, not their causal contributions to retrieval scores.

\subsection{Layer Representations}
\label{app:layer_recall}
\begin{figure}[hbtp]
    \centering
    \includegraphics[width=\linewidth]{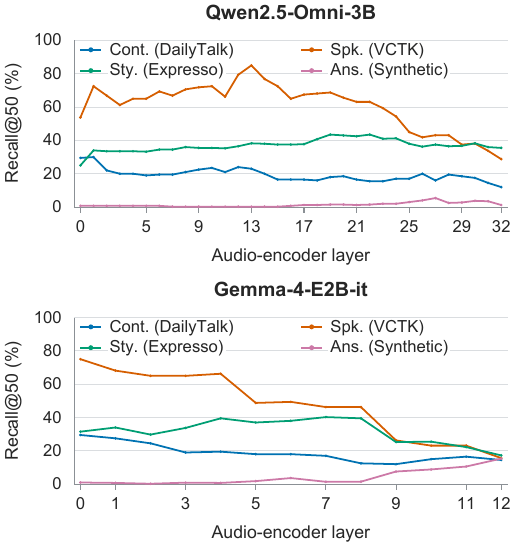}
    \caption{
        Recall@50 using frozen audio-encoder hidden states after audio projection.
        0 denotes the pre-encoder hidden states; encoder layers are indexed from one.
    }
    \label{fig:layer_recall_projected}
\end{figure}

We extend the analysis in Section~\ref{subsubsec:audio_layer_weighting} to examine whether audio projection affects the best-performing layer.
For each layer, we apply the same frozen audio projection, mean-pool the projected hidden states over time, and rank spoken documents by cosine similarity.
Both the audio encoder and projection remain frozen, and the language model is not used.
Figure~\ref{fig:layer_recall_projected} reports Recall@50 for dialogue continuation on DailyTalk, speaker matching on VCTK, speaking-style matching on Expresso, and answer retrieval on Synthetic.
The best-performing layer continues to vary across retrieval requirements after projection.

\end{document}